\documentclass[12pt]{iopart}

\usepackage{cite} 

\usepackage{graphicx}

\usepackage{iopams}
\usepackage{setstack}

\eqnobysec 

\begin{document}

\title[]{Experiments on a videotape atom chip:\\fragmentation and transport studies}

\author{I Llorente Garc\'{i}a, B Darqui\'{e}\footnote{Present address: Laboratoire de Physique des Lasers, CNRS UMR 7538, Universit\'{e} Paris 13, 99 avenue J.-B. Cl\'{e}ment, 93430 Villetaneuse, France.}, E A Curtis\footnote{Present address: National Physical Laboratory, Hampton Road, Teddington, Middlesex TW11 0LW, UK.}, C D J Sinclair\footnote{Present address: University College London Institute of Neurology, Queen Square, London WC1N 3BG, UK.} and E A Hinds}

\address{Centre for Cold Matter, Department of Physics, Imperial College, Prince Consort Road, London SW7 2BW, UK.}
\ead{isabel.llorente-garcia@imperial.ac.uk}

\date{\today}

\begin{abstract}
This paper reports on experiments with ultra-cold rubidium atoms confined in
microscopic magnetic traps created using a piece of
periodically-magnetized videotape mounted on an atom chip. The roughness of the confining potential is studied with atomic clouds at temperatures of a few microKelvin and at distances between
$30\,\mu\mathrm{m}$ and $80\,\mu\mathrm{m}$ from the videotape-chip surface. The inhomogeneities in the magnetic field created by
the magnetized videotape close to the central region of the chip are characterized in this way. In addition, we demonstrate a novel transport mechanism whereby we convey cold atoms confined in arrays of videotape magnetic micro-traps over distances as large as $\sim1\,\mathrm{cm}$ parallel to the chip surface. This conveying mechanism enables us to survey the surface of the chip and observe potential-roughness effects across different regions.
\end{abstract}


\vspace{2pc}


\maketitle

\section{\label{sec:Intro} Introduction}

The use of atom chips to trap neutral atoms and cool them down to temperatures of the order of $10^{-7}$ Kelvin has opened a wide range of possibilities for fundamental studies and applications requiring a high degree of control of an atomic system. An atom chip is a centimetre-scale device that integrates
wires or permanent magnets (or both) in a flat geometry to generate the large magnetic field gradients required to confine neutral atoms at distances of the order of a few to a few hundred micrometres from the surface of the atom chip. Micro-fabricated magnetic traps for neutral atoms were first experimentally realized by J. Reichel \textit{et al.} \cite{chips9} in 1999, and by R. Folman \textit{et al.} \cite{t15} in 2000, using surface-mounted wires on an atom chip. Atom chips offer the advantage of flexible trap configurations that range from single traps to arrays of multiple traps, as well as the possibility of integrating all elements necessary for trapping, cooling, transporting and detecting atoms into a single compact geometry.

Atom chips have been used for a wide variety of fundamental studies, for example: to investigate the behaviour of Bose-Einstein condensates as coherent matter waves in experiments such as atomic beam splitters or atomic interferometers \cite{intf1,intf3,intf5}; to study low-dimensional quantum gases \cite{1DonChip1,1DonChip2,1DonChip3,1DonChip4,intf6}; to explore atom-surface interactions \cite{JonesSpinFlips,CasimirPolder3,ThermalSpinFlips}, etc. Other applications of atom chips include for instance the use of ultra-cold atoms as magnetic field sensors \cite{f14,f16,f26,VortexSensor} with high field sensitivity and high spatial resolution, and the experimental realisation of miniaturized atomic clocks \cite{AtomicClockChip1,AtomicClockChipFernando} which could be incorporated into Global Positioning Systems (GPS) and navigation systems. Atom chips also offer a promising
experimental approach towards the implementation of quantum
information processing with neutral atoms \cite{qi2,qi3,qi4}, since they can combine the accurate control of an atomic quantum system with coherent manipulation and integrated detection tools \cite{Detection5,Detection11,DetectionCavity4,Detection10,Detection8,DetectionCavity5,pyramids,waveguidesPaper} in a miniaturized device with good prospects for scalability. More details can be found in the atom chip review of reference \cite{chipsReview3}, and references therein.\\

The work described in this paper has been carried out with neutral $^{87}$Rb atoms on a permanent-magnet atom chip made of videotape. Several cold-atom experiments around the world are working on atom chips based on different types of permanent magnets such as magnetic films \cite{PermanentMagnets1,PermanentMagnets3}, planar structures of hard magnetic material \cite{PermanentMagnets0,PermanentMagnets6,PermanentMagnets7},
hard disk platters \cite{f22} or ferrimagnetic transparent films
\cite{PermanentMagnets4,PermanentMagnets5}. Magnetic data-storage media such as floppy disk, audiotape and
videotape have been previously investigated in our group and used
to manipulate ultra-cold atoms
\cite{chips5,videofieldcalculation,chips7,chips8,t16,t22}. Several permanently-magnetized atom-chip designs have been investigated theoretically \cite{PermanentMagnets2} as well as from a technical point of view \cite{mofilms}.

The use of videotape to build an atom chip offers the advantage of
miniaturized magnetic patterns that generate strong magnetic
fields (up to $\sim110\,\mathrm{G}$ at the chip surface) and strong field gradients (a few $\mathrm{G}/\mu\mathrm{m}$). The pattern of permanent magnetisation can be chosen according to the needs of the experiment. In our case, the recorded magnetisation is sinusoidal with a wavelength of $106\,\mu\mathrm{m}$, allowing the confinement of rubidium atoms with temperatures between a few hundred $\mu$K and a few hundred nK, in arrays of miniature magnetic micro-traps. An advantage of videotape is the fact that it is made of insulating material. This results in long trapping lifetimes, even close to the videotape \cite{ChrisPRA}, due to the low rate of thermally-induced spin flips compared to the rate near bulk metallic materials \cite{ThermalSpinFlips}.\\

This paper begins by describing the videotape atom chip and the properties of the magnetic micro-traps in section \ref{sec:TheChip}, and continues in section \ref{sec:ExperimentalSequence} by detailing the experimental sequence followed to confine ultra-cold rubidium atoms in these traps. We then focus on two particular subjects. Section \ref{sec:Fragmentation} describes a study of roughness of the trapping potential and fragmentation of atom clouds trapped in very close proximity to the surface of the atom chip. Section \ref{sec:Transport} describes an effective method of transporting cold atoms in videotape magnetic traps over large distances of the order of 1 cm, as a tool to increase the degree of control over the trapped atoms and to survey the surface of our videotape atom chip. More details about previous work on these two subjects are given at the beginning of sections \ref{sec:Fragmentation} and \ref{sec:Transport}.

\section{\label{sec:TheChip} The Chip}

\begin{figure}[ht]
\begin{center}
\includegraphics[width=0.8\textwidth,height=0.4\textwidth]{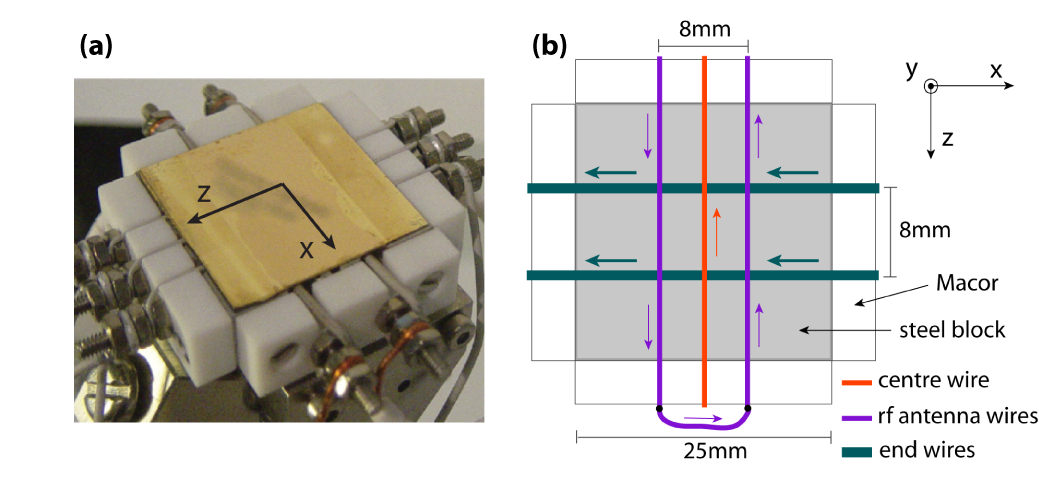}
\end{center}
\caption{\label{fig:TheChip:ChipPhotos} \textbf{(a)}
Photograph of the videotape atom chip. \textbf{(b)} Schematic
view of the wires under the chip. The arrows indicate the direction
of the currents in the wires.}
\end{figure}

Design and fabrication of the videotape atom chip have been described
in detail in our previous papers \cite{ChrisPRA,ChrisEPJD}. A piece
of commercial Ampex 398 Betacam SP videotape with dimensions of 22 mm $\times$ 12.5 mm, was glued onto a glass coverslip and coated with a
400 nm-thick reflective layer of gold. This was subsequently glued onto a block of stainless steel which houses auxiliary wires. Figure \ref{fig:TheChip:ChipPhotos} shows a
photograph of the chip together with a schematic view of the wires
under the chip. The centre wire is used in an intermediate loading stage, together with a uniform bias field along \emph{x}, to create a wire magnetic trap that transfers the atoms from a magneto-optical trap (MOT) into videotape magnetic micro-traps. The radio-frequency (rf)-antenna wires are used for evaporative cooling of the atoms in the magnetic traps and the end wires provide axial confinement. The overall diameter of the centre wire and rf-antenna wires is 0.5 mm, while that of the end wires is 1 mm. All wires are ceramic-coated copper conductors and can sustain currents of up to 20 A during times of up to 20 seconds.


The chip is mounted inside the chamber with its
gold-coated face pointing downwards. The \emph{x}, \emph{y} and
\emph{z} directions shown in figure \ref{fig:TheChip:ChipPhotos}
are used throughout this paper. We establish here that
\emph{y} points down along the vertical direction, with $y=0$
corresponding to the chip surface. The centre wire corresponds to the axis $x=0$, $y=-440\,\mu\mathrm{m}$, and the end wires are located at $y=-1.4\,\mathrm{mm}$, $z=\pm 4\,\mathrm{mm}$, with $z=0$ corresponding to the centre of the chip, half way between the two end wires. Cold atoms are confined below the chip in elongated, tube-like traps, in such a way that \emph{z} corresponds to the axial direction of the traps and \emph{x} and \emph{y}, to the transverse directions.\\

\begin{figure}[ht]
\begin{center}
\includegraphics[width=0.3\textwidth,height=0.26\textwidth]{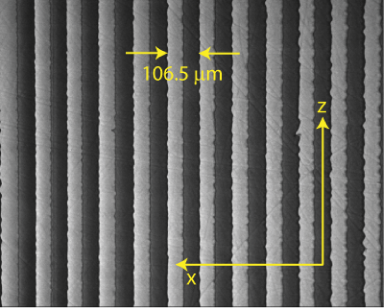}
\end{center}
\caption{\label{fig:TheChip:GarnetImage} Image of the
videotape magnetisation pattern acquired with a polarisation
microscope and a magneto-optical garnet sensor. Note that the noise on the edges of the lines along the \emph{z} direction is due to inhomogeneities in the garnet sensor: displacing the sensor along a fixed spot on the videotape changes the shape of this noise.}
\end{figure}

The videotape is recorded with a periodic pattern of
magnetisation, essentially sinusoidal, given by $\vec{M}=M_1 \cos(kx) \mathbf{\hat{x}}$,
where $k=2 \pi/\lambda$ and $\lambda$ is the spatial period. A detailed calculation of the magnetic field generated by a periodically magnetized videotape can be found in reference \cite{videofieldcalculation}. Figure \ref{fig:TheChip:GarnetImage} shows an image of the videotape magnetisation seen by using a polarisation microscope, with a GGG (Gadolinium Gallium Garnet) magneto-optical sensor \cite{Doetsch} placed on top of a videotape sample. This image is used to measure the period of the recorded pattern, $\lambda = (106.5 \pm 0.4)\,\mu \mathrm{m}$ \cite{MyThesis}.



\begin{figure}[ht]
\begin{center}
\includegraphics[width=0.5\textwidth,height=0.4\textwidth]{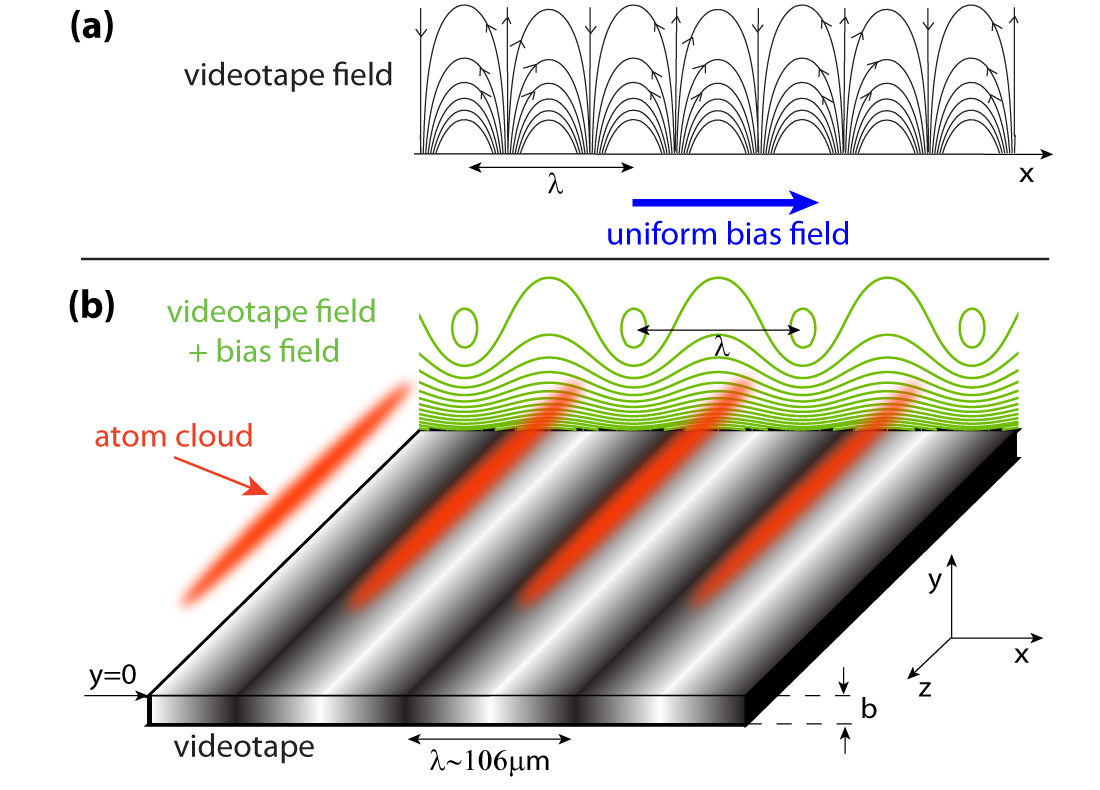}
\end{center}
\caption{\label{fig:TheChip:VideoFieldPlots} Magnetic fields
used to create an array of videotape magnetic micro-traps.
\textbf{(a)} Field lines generated by the sinusoidally magnetized
videotape. \textbf{(b)} Contours of
constant magnetic field strength of combined videotape field and bias field. As a result, atoms can be trapped in an array of elongated magnetic traps separated by a distance $\lambda$ along \emph{x}.}
\end{figure}

The videotape magnetisation generates a periodic magnetic field:
\begin{equation}\label{eqn:videoField}
\vec{B}_{video}=B_1 e^{-ky} \left( -\cos(kx)\, \mathbf{\hat{x}} +
\sin(kx)\, \mathbf{\hat{y}} \right)\, ,
\end{equation}
with
\begin{equation}\label{eqn:videoField2}
B_1=\frac{\mu_0 M_1}{2}(1-e^{-kb})\,,
\end{equation}
where $\mu_0$ is the free-space magnetic permeability and $b$ is the thickness of the magnetized layer in the videotape. The field strength, $B_1$, at the videotape surface is $B_1 =(110 \pm 10)\,\mathrm{G}$  \cite{ChrisEPJD}. Equation (\ref{eqn:videoField}) shows how the videotape field strength decreases exponentially with the distance from the surface. Moving along the \emph{x} coordinate, the field direction rotates while its modulus remains constant. The corresponding videotape magnetic field lines are shown in figure \ref{fig:TheChip:VideoFieldPlots} (a).


When a uniform bias field, $\vec{B}_b=B_b \mathbf{\hat{x}}$, is added along \emph{x}, an array of magnetic micro-guides forms at a distance $y_0$ from the chip surface. This is the height at which the bias field cancels the videotape magnetic field:
\begin{equation}\label{eqn:videoTrapHeight}
y_0=-\frac{1}{k} \ln(\frac{B_b}{B_1}) \, .
\end{equation}
Typical values of $y_0$ range between $20\,\mu \mathrm{m}$ and $120\,\mu \mathrm{m}$. The micro-guides in the array have their axes along \emph{z} and are spaced along \emph{x} by a distance $\lambda \sim 106\, \mu \mathrm{m}$. Figure \ref{fig:TheChip:VideoFieldPlots} (b) shows the contours of constant magnetic field strength that result from combining the videotape field and bias field. Ultra-cold atoms are confined in lines along \emph{z}, at the centre of each closed contour.

\begin{figure}[ht]
\begin{center}
\includegraphics[width=0.5\textwidth]{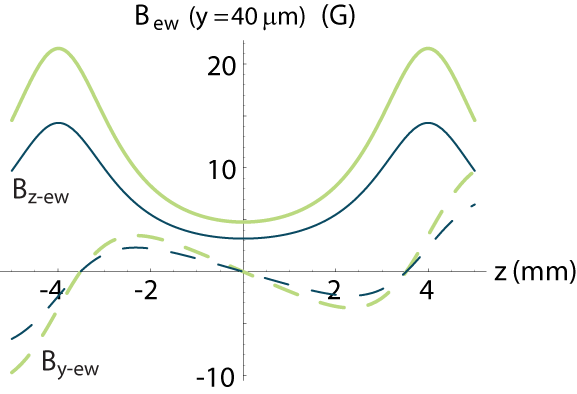}
\end{center}
\caption{\label{fig:EWfields} \emph{Z}-component (solid lines) and \emph{y}-component (dashed lines) of the end-wire field at a height $y=40\,\mu \mathrm{m}$, as a function of \emph{z}. The end wires are located at $z=\pm 4\,\mathrm{mm}$ and $y_0=-1.4\,\mathrm{mm}$. Dark blue lines correspond to a current $I_{end}=10\,\mathrm{A}$, and light green lines to $I_{end}=15\,\mathrm{A}$.}
\end{figure}

Axial confinement is provided by the magnetic field of the two end wires (figure \ref{fig:TheChip:ChipPhotos}). This is of the form $\vec{B}_{ew}(y,z)=B_{y-ew}(y,z)\mathbf{\hat{y}}+B_{z-ew}(y,z)\mathbf{\hat{z}}$. The components $B_{y-ew}$ and $B_{z-ew}$,
evaluated at a height $y=40\,\mu \mathrm{m}$, are plotted in figure \ref{fig:EWfields} as a function of the
axial coordinate of the trap, \emph{z}. Typically, a current $I_{end}=10-15\,\mathrm{A}$ is run through the end wires, leading to axial trap frequencies of $12-15\, \mathrm{Hz}$ at distances up to a few hundred $\mu \mathrm{m}$ from the chip surface.


Helmholtz coils outside the chamber create an additional uniform field along \emph{z}, referred to as $B_{z-coil}$ throughout this paper, which reduces the magnitude of the net field ($B_{z}=B_{z-ew}-B_{z-coil}$) on the axis of the trap.

Close to the central region of the trap, the frequency of small transverse oscillations of the atoms can be expressed as:
\begin{equation}\label{eqn:radialVideofreq}
f_{r}\approx \frac{k B_b}{2\pi}\sqrt{\frac{\mu_B g_F m_F}{m
B_{z}}} = \frac{k B_b}{2\pi}\sqrt{\frac{\mu}{m
B_{z}}}\, ,
\end{equation}
where $m$ is the mass of the atom and $\mu_B g_F m_F$ is the usual factor in the Zeeman energy, which we henceforth abbreviate to $\mu$. The typical transverse oscillation frequencies in the videotape traps range from $500\, \mathrm{Hz}$ to $15\, \mathrm{kHz}$ for bias fields up to $40 \, \mathrm{G}$.

It is possible, by modifying the bias field, the end-wire current and the magnitude of $B_{z-coil}$, to achieve independent control of all the relevant videotape trap parameters, i.e., the distance to the chip surface, the transverse trap frequencies, and the axial trap depth and frequency.


\begin{figure}[ht]
\begin{center}
\includegraphics[width=0.6\textwidth]{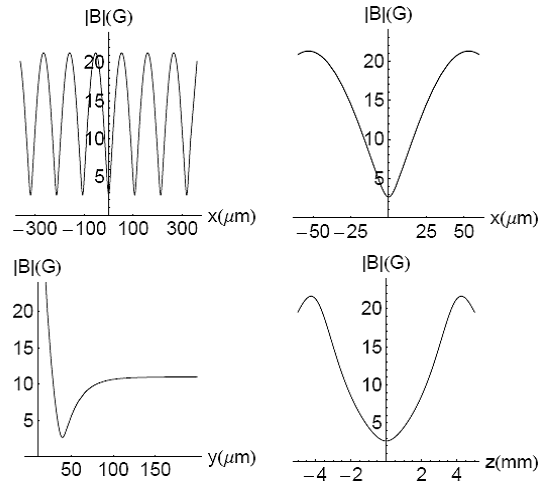}
\end{center}
\caption{\label{fig:VideoFieldsXYZ} Total magnetic
field strength in the videotape magnetic traps as a function of the
\emph{x}, \emph{y}, \emph{z} coordinates (see text for details), for $I_{end}=15\,\mathrm{A}$, $B_b=10.6\,\mathrm{G}$ and $B_{z-coil}=2.6\,\mathrm{G}$. The trap height is $y_0\sim40\,\mu\mathrm{m}$.}
\end{figure}

Figure \ref{fig:VideoFieldsXYZ} plots the modulus of the
total magnetic field, as a function of each coordinate, for typical experimental parameters: $I_{end}=15\,\mathrm{A}$, $B_b=10.6\,\mathrm{G}$ and $B_{z-coil}=2.6\,\mathrm{G}$, which result in a trap-surface separation $y_0\sim40\,\mu\mathrm{m}$. The top plots show $|\vec{B}|$ as a function of
\emph{x}, at $y=y_0$ and $z=0$: the plot on the left-hand side shows
how the periodic potential forms an array of micro-traps separated
by $\lambda \sim 106\,\mu \mathrm{m}$ along \emph{x}, while the plot on the
right zooms into a single micro-trap. The bottom plots
show the dependence of the confining field strength on \emph{y} (left),
evaluating $|\vec{B}|$ at $x=0$ and $z=0$, and on \emph{z} (right),
evaluating $|\vec{B}|$ at $x=0$ and $y=y_0$. Note how the length
scale along \emph{z} is some two orders of magnitude larger than
that along \emph{x} or \emph{y}. Thus, the traps
are very elongated, with their axis along \emph{z}, and can have
aspect ratios up to 1000. The plots also show how the confinement is harmonic only in a limited region close to the centre of the traps.\\

Note that the addition of the end-wire field causes the height of the videotape traps to differ slightly (by $<1\mu\mathrm{m}$) from the expression in equation (\ref{eqn:videoTrapHeight}) (because $B_{z-ew}$ depends on y). Furthermore, $B_{y-ew}$ causes the line of minimum magnetic field strength of the traps, i.e., the effective trap axis, to be curved in three dimensions, instead of being a straight line parallel to \emph{z}. These effects will be discussed in more detail in section \ref{sec:Transport:Mechanism}. As a consequence, the transverse trap frequency slightly differs from that in equation (\ref{eqn:radialVideofreq}), the principal axes of the trap deviate slightly from the \emph{x}, \emph{y} and \emph{z} axes, and thus figure \ref{fig:VideoFieldsXYZ} does not rigorously represent the exact videotape trap cross-sections.

\section{\label{sec:ExperimentalSequence} The experimental sequence}

The experimental sequence that leads to optimum loading of cold atoms into videotape magnetic micro-traps is described here.

A two-chamber set up is used in which a slow beam of atoms ($\leq10\,\mathrm{m/s}$) from a low-velocity intense source (LVIS) \cite{LVIS} loads a reflection MOT at the centre of the main chamber. A rubidium dispenser is turned on for $30\,\mathrm{s}$ and $10^8$-$10^9$ atoms
are captured in the MOT at a distance of $6-7\,\mathrm{mm}$ from the
videotape-chip surface. The MOT temperature is a few hundred $\mu
K$, its density around $10^9-10^{10}\,\mathrm{atoms/cm}^{3}$, and its
lifetime, $50-160\, \mathrm{s}$. The MOT is then moved over $100\, \mathrm{ms}$ to a position $2\, \mathrm{mm}$ away from the chip surface by ramping up an external bias
field of $3.6\, \mathrm{G}$ on the \emph{x}-\emph{y} plane. At this point, the
red-detuning of the MOT beams is increased from $-12\, \mathrm{MHz}$ to $-45\, \mathrm{MHz}$
over $5\, \mathrm{ms}$ and held for another $5\, \mathrm{ms}$, in order to
reduce the temperature of the atoms to $30-60\, \mu \mathrm{K}$ in a
sub-Doppler cooling stage. The trapping light and quadrupole field
of the MOT are then switched off and a $3.7\, \mathrm{G}$ field is turned on
along the \emph{z} direction to act as quantisation axis during an
optical pumping stage. The pumping beam is tuned to the $5\,
^2 S_{1/2}\, F=2 \, \longrightarrow \, 5\, ^2 P_{3/2}\, F=2$
transition in $^{87}$Rb and is circularly polarized ($\sigma^+$
with respect to the quantisation axis). A $400\, \mu \mathrm{s}$-long pulse
of this light pumps the atoms to the $5 \, ^2 S_{1/2}\, F=2 , m_F=+2$ state used for magnetic trapping. Some $5\times10^7$ atoms are then re-captured in a wire
magnetic trap, at a distance of $\sim1.8\, \mathrm{mm}$ from the chip surface,
by suddenly switching on $15\, \mathrm{A}$ through the centre
wire, $15\, \mathrm{A}$ through the end wires, a $17\, \mathrm{G}$ bias field $B_b$ and a $3\, \mathrm{G}$ $B_{z-coil}$. The frequencies of transverse and longitudinal confinement in this trap are $50 \, \mathrm{Hz}$ and $15\, \mathrm{Hz}$, respectively, and the trap depth is around $7\, \mathrm{G}$.
This trap is then adiabatically compressed and moved to a distance
of $200\, \mu \mathrm{m}$ from the chip surface by ramping the bias field up to
$44\, \mathrm{G}$ over $400\, \mathrm{ms}$. This enables us to achieve the high elastic
collision rates required for evaporative cooling. The transverse and
longitudinal trap frequencies in this compressed trap are 500 Hz
and 16 Hz, respectively, resulting in magnetically confined clouds
with aspect ratios of $\sim30$. The temperature of the cloud is
a few hundred $\mu$K at this stage. Two rf evaporation stages are
carried out. First, the
rf is ramped exponentially from $30\, \mathrm{MHz}$ to $11\, \mathrm{MHz}$,
over $3.4\, \mathrm{s}$, with a time constant of $3\, \mathrm{s}$; in the second stage, an
exponential sweep from $11\, \mathrm{MHz}$ to a variable final rf frequency
between $8\, \mathrm{MHz}$ and $2\, \mathrm{MHz}$ is carried out over several seconds, with
a time constant of $5\, \mathrm{s}$. We typically evaporate down to
$1-70\, \mu \mathrm{K}$, with $10^5$-$10^6$ atoms remaining in the trap, and lifetimes of up to $30\, \mathrm{s}$. Curiously, the ionisation gauge controller has to be switched off during the
evaporation to eliminate the noise it generates at these radio
frequencies.

\begin{figure}[ht]
\begin{center}
\includegraphics[width=0.45\textwidth,height=0.47\textwidth]{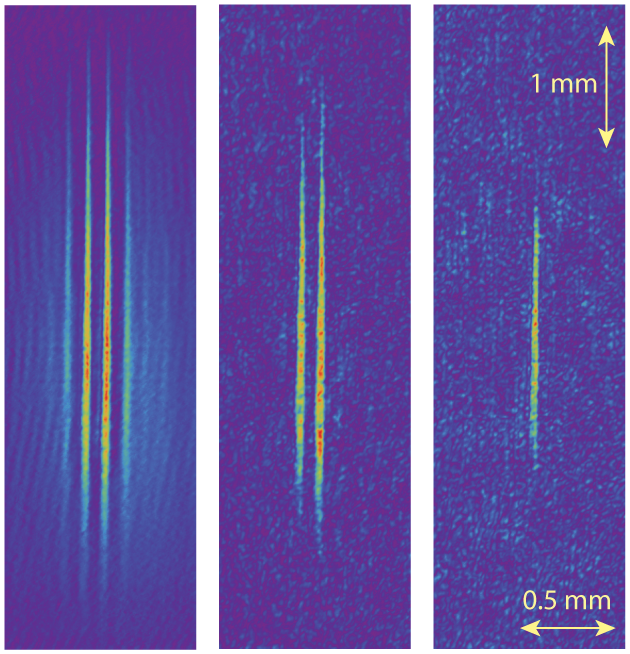}
\end{center}
\caption{\label{fig:LoadOneOrMoreVideoTraps}
In-trap absorption images of atoms loaded into either one or several
videotape traps, acquired using imaging set-up \emph{B} (see text). From left to right, the final rf frequency of the
evaporation ramp in the compressed wire magnetic trap decreases from
$8\, \mathrm{MHz}$ to $2.5\, \mathrm{MHz}$, and the temperature of the best-loaded (central) videotape trap decreases from $\sim50\,\mu\mathrm{K}$ to $\sim10\,\mu\mathrm{K}$. The length scale is the same for all images.}
\end{figure}

The cooled atoms are transferred from the wire trap into either one or several videotape magnetic traps in two stages, with durations of $1\, \mathrm{s}$ and $300\, \mathrm{ms}$, respectively. In the first transfer stage the
centre-wire current is lowered from $15\, \mathrm{A}$ to $6\, \mathrm{A}$, the horizontal
bias field is decreased from $44\,\mathrm{G}$ to $22\,\mathrm{G}$, and the axial field is
decreased from $2.8\,\mathrm{G}$ to $0\,\mathrm{G}$. In the second transfer stage the
currents are lowered to $0\, \mathrm{A}$ in the centre wire and $10\, \mathrm{A}$ in the end wires, and the bias field is decreased to $2.2\,\mathrm{G}$, leaving
the atoms confined in videotape traps $\sim 70$
$\mu$m from the chip surface. The temperature of the atoms after evaporation controls the number of videotape traps loaded. A $50\, \mu \mathrm{K}$-cloud loads 4 videotape traps at the same temperature with
approximately $8\times 10^5\,\mathrm{atoms}$  in the fullest (central) trap, while a $10\, \mu\mathrm{K}$-cloud in the wire trap loads a single $10\, \mu\mathrm{K}$ videotape trap with some $2\times
10^5$ atoms (see figure \ref{fig:LoadOneOrMoreVideoTraps}).

Atoms in these videotape traps can then be moved to distances from
the chip surface between $20\, \mu \mathrm{m}$ and $100\, \mu \mathrm{m}$ by ramping the
bias field to a value between $0.5\,\mathrm{G}$ and $30\,\mathrm{G}$ (typically over $1\, \mathrm{s}$).
The longitudinal trap frequency is $\sim 13\, \mathrm{Hz}$, and the transverse
trap frequencies range between $0.5\, \mathrm{kHz}$ and $15\, \mathrm{kHz}$, leading to
aspect ratios between 40 and 1000. An additional rf evaporation
stage can cool
the atoms down to around $500\, \mathrm{nK}$ if needed. Lifetimes in excess of $30\, \mathrm{s}$ are measured for cold atoms ($\sim20\,\mu\mathrm{K}$) confined in videotape
traps $\sim 70\, \mu \mathrm{m}$ away from the chip surface.


At the end of the experimental sequence, an absorption image of the
atoms, either in-trap or after
release, is recorded using a resonant imaging beam and a CCD
camera. Two imaging set-ups with unit magnification are in place.
In set-up \emph{A}, an imaging beam reflects off the
gold-coated surface of the chip at an angle of $14^{\circ}$ and
propagates at $35^{\circ}$ to the \emph{x} direction.
A double absorption image appears, corresponding to the atomic
cloud and its reflection in the chip surface. The atom-surface
separation is half the perpendicular distance between these two
images. The imaging beam in set-up \emph{B} propagates in the
\emph{y}-\emph{z} plane, reflecting off the chip surface at
$45^{\circ}$. The cloud and its reflection are unresolved and form a single image, as shown in figure \ref{fig:LoadOneOrMoreVideoTraps}. The oblique viewing is responsible for the different \emph{x} and \emph{z} scales in this figure. The spatial resolution is $\sim7 \,\mu \mathrm{m}$ for both arrangements.

\section{\label{sec:Fragmentation} Fragmentation experiments}

A thermal cloud of atoms confined in a corrugated magnetic potential will start breaking into fragments when the thermal energy of the atoms is comparable to the depth of the corrugation, which on a typical atom chip is a few $\mu$K for atom-surface separations below $100\,\mu \mathrm{m}$. For a Bose-Einstein condensate (BEC) the relevant energy scale would be the chemical potential instead.

In some cases, this corrugation can turn out to help experiments by, for instance, resulting in tighter confinement \cite{interferometerSewell} or leading to the production of multiple BECs \cite{f21}. However, it is worth noting that, in such cases, the landscape of the corrugation is uncontrolled. In other cases, where a smooth trapping potential is required, fragmentation effects are clearly undesirable.

Fragmentation of atom clouds near current-carrying wires, as first reported in references \cite{f1,f2,f3,f4,f7,f6}, is due to a magnetic field component along the wire. This anomalous component, typically several orders of magnitude smaller than the main transverse component, is due to deviations
in the direction of the current flow. Further detailed studies of fragmentation near conducting wires have been reported in
references
\cite{f5,f10,f8,f9b,f11,f17,f36,f14,f16,f26}. In most
of these chips, imperfections in the wires, such as edge
roughness or bulk defects, were found to be responsible. Substantial improvements in the techniques for micro-fabricating wires on atom chips have led to smoother trapping potentials in the past few years.

Potential roughness has also been observed in atom chips based on
permanently magnetized materials, such as
videotape \cite{ChrisEPJD}, magneto-optical films
\cite{f19} or hard disk \cite{f22}. Potential roughness in those cases was attributed to inhomogeneity of the magnetic material.\\

This section presents a study of axial fragmentation in our videotape magnetic traps. Absorption images of fragmented clouds are analysed to obtain the axial disorder potential at several distances from the videotape surface. The dependence of the rms potential roughness on atom-surface separation is studied and the strength of this roughness is compared to that observed above other types of atom chip. The frequency spectrum of the disorder potential is analysed and the possible origins of fragmentation on our atom chip are discussed. Finally, a method for reducing fragmentation is outlined and discussed in the context of our videotape atom traps.

\subsection{\label{sec:Fragmentation:first} Fragmentation above the videotape atom chip}

As described in section \ref{sec:TheChip}, the periodic, in-plane
magnetisation recorded in the videotape along the \emph{x} direction
should generate a magnetic field with components only in the
\emph{x}-\emph{y} plane (see equation (\ref{eqn:videoField})). However, we observe fragmentation
due to a fluctuating anomalous magnetic field component, $\Delta
B_z$, along the axis of the cloud (\emph{z}). Note that only the \emph{z}-component of the noise in the trapping field can lead to axial fragmentation. Noise in the other components has a negligible effect since it only causes a slight transverse displacement of the trap centre.

The disorder potential and the anomalous field component, $\Delta
B_z$, are obtained by measuring the density profile of an ultra-cold atomic cloud using absorption imaging. For a cloud of atoms in thermal equilibrium at temperature $T$, the linear density profile, $n(z)$, is proportional to the Maxwell-Boltzmann probability distribution, so that we can write $n(z)=b \exp[-(U_{dis}(z)+U_{trap}(z))/(k_B T)]$, where $b$ is a normalisation constant, $k_B$ is Boltzmann's constant, $U_{dis}(z)$ is the disorder potential and $U_{trap}(z)$ is the intended axial potential. The disorder potential can therefore be expressed as:
\begin{equation}\label{eqn:Fragm:disorderPot}
U_{dis}(z)=-k_B T \ln \left( \frac{n(z)}{b} \right)-U_{trap}(z).
\end{equation}
The axial potential created by the end wires is
approximately harmonic in the regions of the trap explored by the
atoms, so we can write $U_{trap}(z)\simeq
\frac{1}{2}m \omega_z^2(z-z_0)^2$, where $m$ is the mass of $^{87}$Rb, $\omega_z$ is the axial oscillation frequency and $z_0$ is the central position of the cloud.

The ordered component of the
potential, $U_{trap}(z)$, is obtained first, by fitting the measured
linear density to a gaussian
function, $b \exp \left[-(\frac{1}{2}m
\omega_z^2(z-z_0)^2)/(k_B T)\right]$, plus a background. The disorder potential is then found by subtracting the smooth harmonic trapping potential
from the full potential measured with atoms.
On the trap axis, where the transverse magnetic field is zero,
the disorder potential is related to the anomalous longitudinal
magnetic field by:
\begin{equation}\label{eqn:Fragm:disorderPot3}
U_{dis}(z)=\mu \Delta B_z(z).
\end{equation}

\begin{figure}[ht]
\begin{center}
\includegraphics[width=0.5\textwidth]{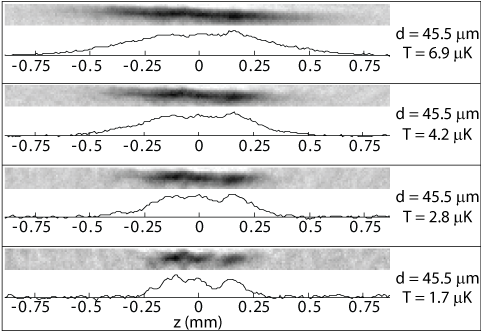}
\end{center}
\caption{\label{fig:Fragmentation:FragmVsTemp} Fragmentation
in a videotape atom trap situated a distance $\mathrm{d}=45.5\,\mu\mathrm{m}$ away from the chip surface. Absorption images and axial linear density profiles for different cloud temperatures, T.}
\end{figure}

\begin{figure}[ht]
\begin{center}
\includegraphics[width=0.4\textwidth]{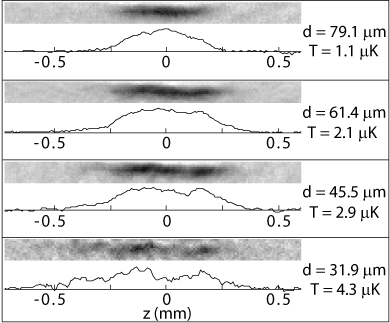}
\end{center}
\caption{\label{fig:Fragmentation:FragmVsHeight}
Fragmentation in a videotape atom trap at different distances, $d$, from the chip surface. The absorption images and axial linear density profiles show how the corrugations in the trapping potential are more pronounced for small atom-surface separations.}
\end{figure}

We have studied atoms with temperatures of a few $\mu$K at distances between $32\,\mu\mathrm{m}$ and $79\,\mu\mathrm{m}$ from the chip surface. Absorption images of the trapped atoms, as shown in figures \ref{fig:Fragmentation:FragmVsTemp} and \ref{fig:Fragmentation:FragmVsHeight}, are recorded at the end of the experimental sequence described in section \ref{sec:ExperimentalSequence}, using imaging set up \emph{A}.

Figure \ref{fig:Fragmentation:FragmVsTemp} shows axial density profiles of atom clouds
confined $45.5\,\mu\mathrm{m}$ away from the chip surface. The temperature of
the cloud decreases from top to bottom, illustrating how the fragments become more pronounced as the
temperature of the atoms becomes comparable with the depth of
the disorder potential.


Figure \ref{fig:Fragmentation:FragmVsHeight} shows atomic
clouds with temperatures of the same order of magnitude, around $1-4\,\mu\mathrm{K}$, but at different distances from the chip surface. As the atoms move closer to the surface, we see the corrugations becoming stronger.\\


\begin{figure}[ht]
\begin{center}
\includegraphics[width=0.5\textwidth]{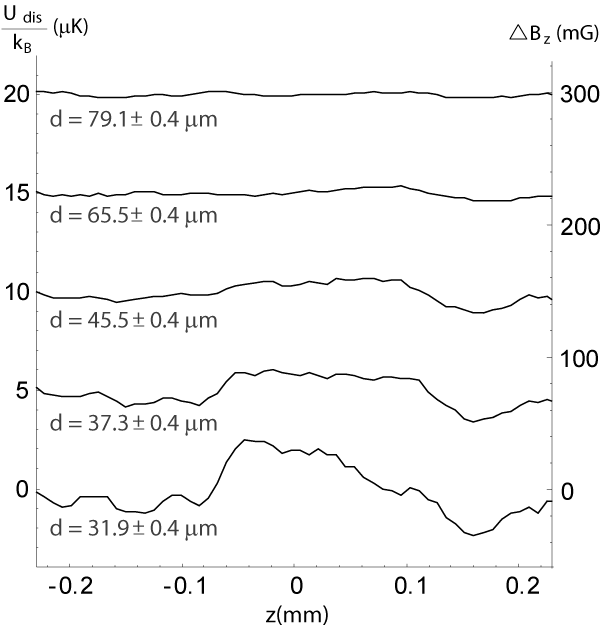}
\end{center}
\caption{\label{fig:Fragmentation:allFragmPots} Axial
disorder potentials measured with ultra-cold atoms in videotape
magnetic traps. Each curve corresponds to a different atom-surface
separation, $d$, that decreases from top to bottom, as indicated by the labels, and is an average of 4-6 experimental realisations. The
potentials have been offset vertically from each other by $5\,\mu\mathrm{K}$ for clarity. The anomalous axial magnetic field, $\Delta B_z$, is calculated using equation (\ref{eqn:Fragm:disorderPot3}).}
\end{figure}

For each atom-surface separation, $d$, we record 4-6 absorption images at a range of temperatures between $1.3\,\mu\mathrm{K}$ and $12.6\,\mu\mathrm{K}$. Each image is used to obtain the axial disorder potential making use of equations (\ref{eqn:Fragm:disorderPot}) and (\ref{eqn:Fragm:disorderPot3}). These potentials are averaged to produce the map shown in figure \ref{fig:Fragmentation:allFragmPots}. This shows how the axial potential roughness decreases away from the surface, while maintaining the same general shape. At a distance of
$32\,\mu\mathrm{m}$, the peak-to-peak amplitude
of the disorder potential is $\sim3\,\mu\mathrm{K}$, decreasing to $\sim0.2\,\mu\mathrm{K}$ at $79\,\mu\mathrm{m}$.\\


\begin{figure}[ht]
\begin{center}
\includegraphics[width=0.6\textwidth]{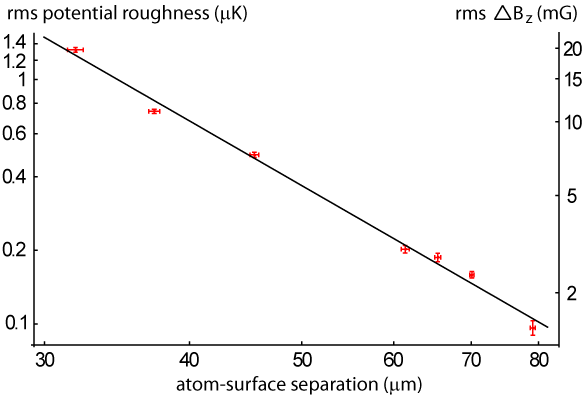}
\end{center}
\caption{\label{fig:Fragmentation:rmsRoughnessVsHeight}
Potential roughness as a function of atom-surface separation $d$ (logarithmic scales). Each data point results from the analysis of 4-6 absorption images coming from different realisations of the experimental sequence. The solid line shows a linear fit of the data that yields a dependence of roughness on atom-surface separation of $d^{-2.7}$.}
\end{figure}

Figure \ref{fig:Fragmentation:rmsRoughnessVsHeight} shows a log-log plot of the measured rms roughness in $\mu \mathrm{K}$ (and the corresponding noise $\Delta B_z$). A linear fit to this graph gives roughness of the form $d^{-B}$, with $B=2.7 \pm 0.1$. The magnification of the imaging set-up was $1.0 \pm 0.1$, introducing a scaling uncertainty of $10\%$ for the horizontal axis in figure \ref{fig:Fragmentation:rmsRoughnessVsHeight}, but this has no effect on the value of $B$.\\


Several experimental groups that use current-carrying wires to confine cold atoms on a chip have measured the decay of roughness with distance $d$. Reference \cite{f5} found an exponential dependence $\exp(-k_0 d)/\sqrt{k_0 d}$, consistent with an oscillating current flow inside the wire with transverse wavevector $k_0$. Two papers \cite{f9b,f11} calculated a dependence of $d^{-2.5}$, valid for distances much larger than the width of the wire, using a model based on white-noise fluctuations of the edges of a flat wire, or of its surface topography. The experimental data presented in references \cite{f8,f11} was consistent with this model. The calculations in reference \cite{f36} took into consideration the self-affine fractal character of the roughness measured in micro-fabricated wires, as opposed to white-noise fluctuations. Cold atoms confined at distances small compared to the width of the wire \cite{f17} showed a decay of potential roughness faster than $d^{-2.5}$. In this case, wire-edge fluctuations were no longer the dominant cause for fragmentation, but instead, inhomogeneous conductivity and top-surface roughness of the wire were more probable explanations.

As for permanent-magnet atom chips, a dependence of $d^{-1.85}$ was measured with cold atoms trapped next to a magneto-optical film \cite{f19}. A calculation that included two-dimensional white-noise spatial variations in the magnetisation component perpendicular to the film led to a similar power law of $d^{-2}$.\\

Table \ref{table:CompareRoughness} compares the roughness measured above several atom chips, which we express in temperature units to allow easy comparison with the temperature of the atom cloud. Wires micro-fabricated using lithographic patterning of evaporated gold \cite{f11,f17} have proved to be substantially smoother than macroscopic wires or other micro-wires. The roughness of the videotape potential is below that of the other atom chips except for reference \cite{f17}.\\

\begin{table}[ht]
\centering
\caption{\label{table:CompareRoughness}Comparison of measured rms potential roughness for different experimental groups working on atom chips.}
\lineup
\begin{tabular}{@{}llll}
\br
\footnotesize{rms roughness}  &  \footnotesize{$d$($\mu$m)} & \footnotesize{Atom chip} & \footnotesize{ref / group} \\
\mr
$\sim$ 2 $\mu$K     & 40  & \footnotesize{Macroscopic wire, 0.5 mm diameter.} & \footnotesize{\cite{f5}/Hinds}\\
$\sim$ 2 $\mu$K     & 33  & \footnotesize{Gold electroplated micro-wire.} & \footnotesize{\cite{f8}/Aspect}\\
a few $\mu$K  & 45  & \footnotesize{Silver foil wires, micro-cut.}  & \footnotesize{\cite{f10}/R.-Dunlop} \\
a few $\mu$K  & \04   & \footnotesize{Micro-wires: lithography $+$ gold evaporation.} & \footnotesize{\cite{f17}/Schmiedmayer} \\
$\sim$ 7 $\mu$K     & 60  & \footnotesize{TbGdFeCo multilayer magnetic film.} & \footnotesize{\cite{f19}/Hannaford}\\
$\sim$ 0.5 $\mu$K   & 45  & \footnotesize{Videotape} & \footnotesize{this paper/Hinds} \\
\br
\end{tabular}
\end{table}

\begin{figure}[ht]
\begin{center}
\includegraphics[width=0.55\textwidth]{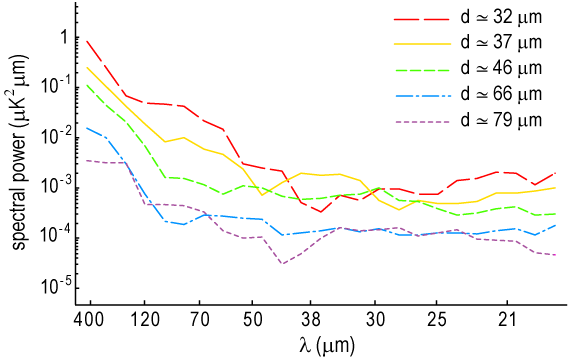}
\end{center}
\caption{\label{fig:Fragmentation:Spectra} Fourier power
spectra of the disorder potentials measured in the videotape traps
for different atom-surface separations, $d$.}
\end{figure}

The Fourier spectra of our disorder potentials are presented in figure \ref{fig:Fragmentation:Spectra} for several atom-surface separations. The power of all frequency components increases at smaller atom-surface separation. At a given value of $d$, the most important contributions are from wavelengths $\gtrsim d$. We do not detect potential variations on a length scale much less than the $7\,\mu\mathrm{m}$ imaging resolution. Similar spectra have been observed in several other atom-chip experiments \cite{f8,f11,f17,f19}.


\subsection{\label{sec:Fragmentation:origin} Origin of fragmentation in the videotape atom chip}

In considering the origin of roughness, we distinguish two effects. (\textit{i}) The tape could be wrinkled or the magnetic layer could have varying thickness. (\textit{ii}) The magnetic material could be inhomogeneous due to density variations or defects.

\begin{figure}[ht]
\begin{center}
\includegraphics[width=0.55\textwidth,height=0.48\textwidth]{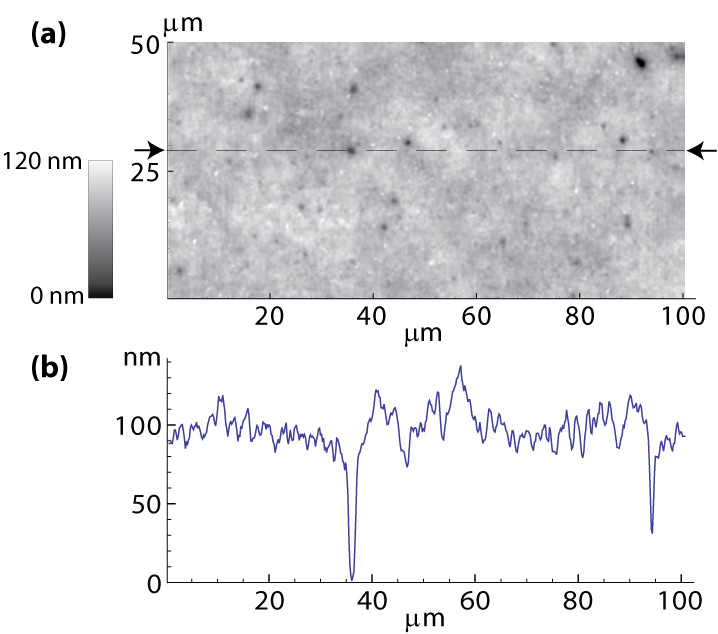}
\end{center}
\caption{\label{fig:Fragmentation:AFM} Atomic force microscope scan showing the topography of a piece of videotape. \textbf{(a)} Survey of a $100 \, \mu \mathrm{m} \times 50 \, \mu \mathrm{m}$ region. \textbf{(b)} Height versus position along the line indicated in (a).}
\end{figure}

The surface topography of the magnetic layer of a piece of videotape
was inspected using an atomic force microscope (AFM), as shown in figure \ref{fig:Fragmentation:AFM}. Height variations of up to $20\,\mathrm{nm}$ were found with wavelengths of the order of tens of microns. These could result from the inability of the videotape to lie completely flat or from variations in its thickness. The AFM scans also revealed some deeper holes in the surface of the
magnetized layer that were typically around $100\,\mathrm{nm}$ deep and a few $\mu$m wide \cite{ChrisEPJD}. The $3.5\,\mu\mathrm{m}$-thick magnetic layer of the videotape consists of iron-composite needles $100\,\mathrm{nm}$ in length and $10\,\mathrm{nm}$ in radius, embedded in glue and aligned parallel to each other along the \emph{x} direction. A hole such as the deep ones
observed on the AFM scans would constitute a magnetisation defect
with hundreds of magnetized needles missing. The magnetic field
created by the videotape directly below one of these defects would
have a non-zero component along
the \emph{z} direction \cite{ChrisThesis}. It seems most probable that the combined effects of height and
thickness variations and defects in the videotape are responsible for the measured fluctuations in the \emph{z}-component of the total magnetic field. The videotape used in our experiment has the smoothest surface topography of several types \footnote{We have inspected the following digital videotapes: Quantegy DBC-D12A, Quantegy D2V-126LC and Fuji D2001-D-2-S-12. Analog videotape is no longer commercially available.} inspected through atomic force microscopy.\\


\subsection{\label{sec:Fragmentation:suppressingFragmentation} Suppressing fragmentation in the videotape atom chip}

The fragmentation described above constitutes an important obstacle to studies of one-dimensional (1D) quantum gases on this chip. Since the large transverse trap frequencies required to enter the 1D regime can only be achieved at small atom-surface separations, an effective mechanism for suppressing fragmentation would prove an extremely useful tool for the progress of such experiments.

As reported in references \cite{f24,st18}, fragmentation can be reduced in a current-carrying wire atom chip by rapid reversal of the currents through the trapping and bias wires. This reverses $\Delta B_z$, so that the time-averaged potential has a reduced roughness. In our experiment, we cannot reverse $\Delta B_z$, but we can instead overwhelm it with a rotating transverse field, as we describe below.

We propose reducing $B_{z}$ to zero, which gives a two-dimensional, transverse quadrupole field. A rotating magnetic field
$B_{rot}$, added in the transverse plane, causes the magnetic field minimum to describe a circle. If the rotation frequency is
high enough compared to the trapping frequencies \footnote{strictly speaking, we should also compare to the actual oscillation frequencies of the atoms inside the fragments, since these may be considerably higher than the nominal trapping frequencies} and low enough
compared to the Larmor frequency, the atoms are unable to follow the
movement of the field minimum and feel a
time-averaged potential. The result is a TOP
(time-averaged orbiting potential) trap \cite{TOPtrap}, in which the time average over one field rotation
yields a parabolic transverse potential close to the trap centre,
with the field at the bottom of the trap being equal to $B_{rot}$:
\begin{equation}\label{eqn:Fragmentation:removeFragm0}
|\vec{B}(r)|\approx B_{rot}+\frac{\alpha^2}{4B_{rot}}r^2\,,
\end{equation}
where $\alpha=k B_b$ is the gradient of the static transverse
quadrupole field. From equation
(\ref{eqn:Fragmentation:removeFragm0}) we deduce that the
frequency of transverse oscillations in the TOP trap is $f_{r-TOP}=\sqrt{B_{z}/(2 B_{rot})} f_r$, where $f_r$ is the frequency in a static trap with axial field $B_z$, given by equation (\ref{eqn:radialVideofreq}).

The transverse field strength at the bottom of the
time-averaged potential is $B_{rot}$, and the \emph{z}-component from the
end wires, assumed to be harmonic, is $\beta z^2$, so that:
\begin{equation}\label{eqn:Fragmentation:removeFragm}
|\vec{B}(r=0,z)|=\sqrt{B_{rot}^2+(\beta z^2)^2}\approx
B_{rot}+\frac{\beta^2}{2B_{rot}}z^4\,.
\end{equation}
This shows that the harmonic axial confinement is now flattened to the more box-like form $z^4$, offering interesting possibilities for studies of one-dimensional quantum gases \cite{TonksReichelBoxLike}.

Adding the noise field, $\Delta B_z$, equation (\ref{eqn:Fragmentation:removeFragm}) becomes:
\begin{eqnarray}\label{eqn:fragmentation1}
 |\vec{B}(r=0,z)|=\sqrt{B_{rot}^2+\left( \Delta B_z + \beta z^2 \right)^2} \,.
\end{eqnarray}
If we choose $B_{rot}$ to be large compared to $\left( \Delta B_z + \beta z^2 \right)$,
\begin{eqnarray}\label{eqn:fragmentation2}
 |\vec{B}(r=0,z)| \approx B_{rot}+\frac{\beta^2}{2 B_{rot}}z^4+\left( \frac{\Delta B_z}{2 B_{rot}}+\frac{\beta z^2}{B_{rot}} \right) \Delta B_z,
\end{eqnarray}
\noindent{where} the first two terms duplicate equation (\ref{eqn:fragmentation1}) and the last term is the roughness suppressed by the factor $\left( \frac{\Delta B_z}{2 B_{rot}}+\frac{\beta z^2}{B_{rot}} \right)$. Near the bottom of the axial potential this factor is approximately $\Delta B_z/(2 B_{rot})$.


\begin{figure}[ht]
\begin{center}
\includegraphics[width=0.5\textwidth]{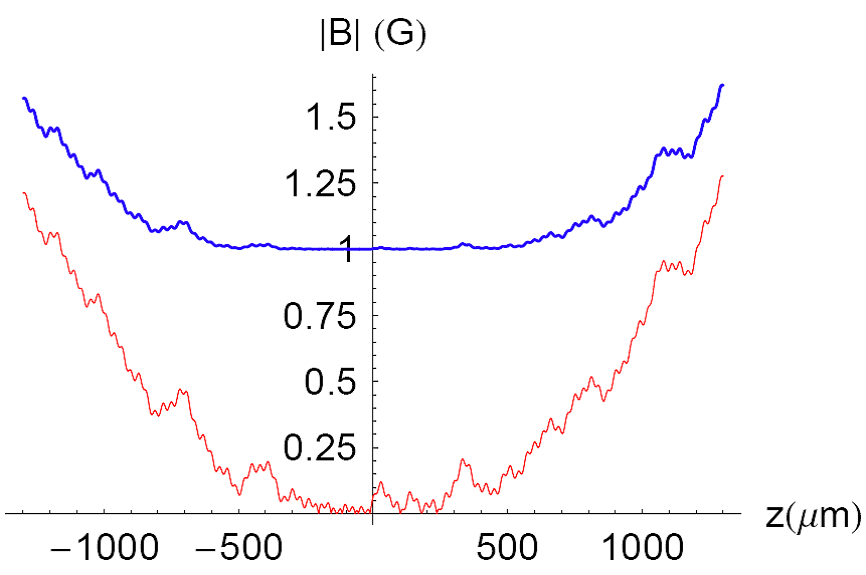}
\end{center}
\caption{\label{fig:Fragmentation:suppress} Predicted
reduction of the axial potential roughness by cancelling
the net axial offset field and superimposing a rotating
transverse field of $1\,\mathrm{G}$. The thin red line shows the static magnetic field strength with disorder, including the end-wire field, before the rotating field is added. The thicker blue line shows the time-averaged field strength after adding the rotating field.}
\end{figure}

Figure \ref{fig:Fragmentation:suppress} shows a calculated example of this smoothing method in a trap $\sim 22\,\mu\mathrm{m}$ from the chip surface ($I_{end}=15\,\mathrm{A}$, $B_b=30\,\mathrm{G}$, $B_{z-coil}=4.73\,\mathrm{G}$ and $B_z=0\,\mathrm{G}$). At this height the rms field corrugation would be of order $50\,\mathrm{mG}$ ($3.3\,\mu \mathrm{K}$). The total magnetic
field modulus is plotted as a function of the position, \emph{z}, along the axis of the trap. The thin red curve shows the field strength before the
rotating field is included. A $1\,\mu \mathrm{K}$ atom cloud would appear strongly fragmented in such a trap. The thicker blue curve shows the resulting
time-averaged field strength after a rotating field, $B_{rot}=1\,\mathrm{G}$, is applied.
The field fluctuations are strongly reduced in the central region of the trap, where $\left( \Delta B_z + \beta z^2 \right) < B_{rot}$, the maximum reduction factor being $1/40$ at $z=0$. The graph also shows how the field at the bottom of the trap is raised by an offset equal to $B_{rot}$, as given by equation (\ref{eqn:fragmentation2}). For the chosen values of $B_b$ and $B_{rot}$, the transverse frequency in the time-averaged potential would be $f_{r-TOP}\sim16\,\mathrm{kHz}$, of the same order of magnitude as the transverse frequency in the static trap ($23\, \mathrm{kHz}$) with $B_{z}=1\,\mathrm{G}$. Considering this frequency and the Larmor frequency of a few MHz, a transverse rotating field with a frequency of order $50\, \mathrm{kHz}$ would be a good choice. Note that we require $B_{rot}\gg 4 k_B T/\mu_B$, where $T$ is the temperature of the atoms in the trap, in order to prevent the atoms reaching the circle of zero field \cite{TOPtrap}. For a $1\,\mu\mathrm{K}$ cloud, this implies $B_{rot}\gg 0.06\,\mathrm{G}$, which is amply satisfied in the example above.\\

This possibility of maintaining large transverse trap frequencies together with a flatter axial potential and a strong suppression of the potential roughness makes our videotape traps interesting for possible future studies of 1D quantum gases.

\section{\label{sec:Transport} Transport experiments}

The possibility of moving atoms from one region to another promises to enhance the functionality of atom chip devices. For this reason transport has been studied by several groups.

Transport of cold atoms along small magnetic guides on a chip has been reported in references \cite{t15,t19,f4}. The group of T. W. H\"{a}nsch and J. Reichel \cite{t8,t2,t1,t33} realised a magnetic conveyor belt for moving atoms trapped in 3D across a chip over distances of many $\mathrm{cm}$. A similar chip set up which demonstrated adiabatic transport of a $6\,\mu\mathrm{K}$ thermal cloud over several $\mathrm{mm}$ is reported in reference \cite{t10}. The group of R. J. C. Spreeuw \cite{t20} displaced atoms up to a round-trip distance of $360\,\mu\mathrm{m}$, in a two-dimensional array of hundreds of tight traps generated by a permanently magnetised FePt film.

We have now brought similar capability to our chip by implementing the effective transport of cold atoms over long distances up to $\sim1\,\mathrm{cm}$ while they remain confined in all three dimensions. This is achieved by rotating the direction of the applied bias field in order to translate the traps along \emph{x}, as proposed in \cite{t16,t22,t18,chips8}. We have used this capability to survey the chip surface in our study of disorder above different regions of the videotape, as discussed in section \ref{sec:Transport:ExperimentalData}.

\subsection{\label{sec:Transport:Mechanism} Transport mechanism}

\begin{figure}[ht]
\begin{center}
\includegraphics[width=0.5\textwidth]{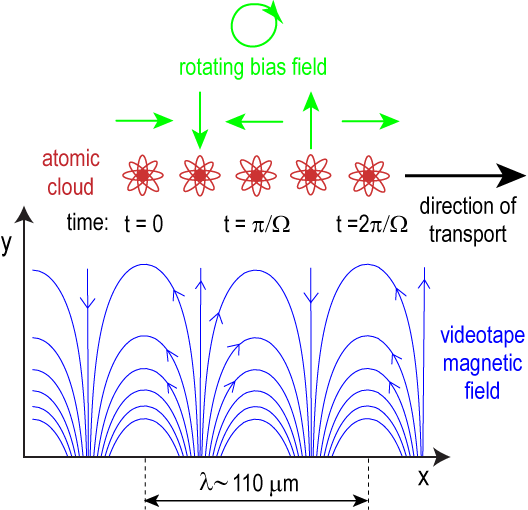}
\end{center}
\caption{\label{fig:Transport:Mechanism} Transport mechanism used to displace the trapping potential along \emph{x}. The videotape magnetic field lines are shown in blue and the rotating bias field in green. As the orientation of the bias field changes in time during one rotation, the traps translate along \emph{x} by a distance equal to $\lambda$. $\Omega/(2 \pi)$ is the bias-field rotation frequency.}
\end{figure}

At a fixed distance from the chip surface, the magnetic field created by the magnetized videotape has constant modulus and its direction rotates as one moves along \emph{x}, as given by equation (\ref{eqn:videoField}). Since the traps are centred on the lines where the bias field cancels the videotape field, a rotation of the bias field translates the traps along the \emph{x}-direction, as illustrated by figure \ref{fig:Transport:Mechanism}. One full rotation translates the traps by one spatial period
$\lambda \sim110\,\mu\mathrm{m}$, giving a translation
velocity $v_x =\lambda \Omega/(2 \pi)$, where $\Omega/(2 \pi)$ is the rotation frequency of the bias field.\\


It is important to consider the contribution of all the magnetic fields that confine the atoms during a transport cycle. In particular, the \emph{y}-component of the field created by the end wires (see figure \ref{fig:EWfields}) changes sign as we move along the \emph{z} axis and cross $z=0$, where its value is zero. This gives rise to a bending of the trap axis and to a variation of the transverse trap frequencies along the trap axis, causing a transverse compression at one end of the trap and de-compression at the opposite end. Both effects depend on the direction of the bias field. As a consequence, two dynamic effects take place during transport: there is an oscillating bend of the trap axis, and a periodic transverse compression and de-compression of the trap ends. In addition there is a small periodic movement of the trap centre. These effects are explained in more detail below.

\begin{figure}[ht]
\begin{center}
\includegraphics[width=0.6\textwidth]{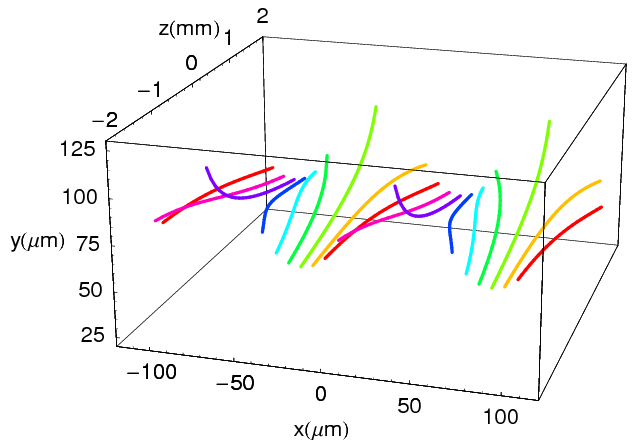}
\end{center}
\caption{\label{fig:Transport:minFieldLine3D} Line of minimum potential energy ($\zeta$) as seen in 3D during
two cycles of bias-field rotation for the transport of atoms in
videotape traps with $\lambda=110\,\mu\mathrm{m}$, $B_b=2.2\,\mathrm{G}$, $B_{z-coil}=0$ and
$I_{end}=10\,\mathrm{A}$, corresponding to our experimental parameters. Gravity is included in the calculation.}
\end{figure}

The line of minimum potential energy of the trap, denoted as $\zeta$ axis, is bent and evolves during transport as shown in figure \ref{fig:Transport:minFieldLine3D}. The line $\zeta$ is calculated numerically by finding the minimum of the potential energy in the \emph{x-y} plane (including gravity) for a given \emph{z} value, then moving along \emph{z} between $z=\pm2\, \mathrm{mm}$. We exaggerate the effect by plotting mm along \emph{z} and $\mu \mathrm{m}$ along the other axes.


When the bias field is along the \emph{x} direction (red and light-blue lines in figure \ref{fig:Transport:minFieldLine3D}, at $x=0,\pm 55\, \mu \mathrm{m}, \pm 110\, \mu \mathrm{m}$), the trap is mostly curved horizontally on the \emph{x-z} plane because the \emph{y}-field from the end wires is up at one end and down at the other. This rotates the net bias field oppositely at the two ends, displacing the trap towards +\emph{x} at one end and -\emph{x} at the other. At the ends of the trap, where $|z|\sim 2\,\mathrm{mm}$, the trap height also decreases by a few micrometres, due to the increase of the total modulus of the bias field caused by the added \emph{y}-component of the end-wire field. When the bias field is along the \emph{y} direction (light-green and purple lines in figure \ref{fig:Transport:minFieldLine3D}, at $x= \pm 27.5 \, \mu \mathrm{m}, \pm 82.5\, \mu \mathrm{m}$), the bending of the trap is vertical, on the \emph{y-z} plane, since the \emph{y}-component of the end-wire field simply adds to the total bias field, increasing its magnitude on one side of $z=0$ and decreasing it on the other. For the two cases described above, the tilt angles near the trap centre are $\sim 1^{\circ}$.


\begin{figure}[ht]
\begin{center}
\includegraphics[width=0.6\textwidth]{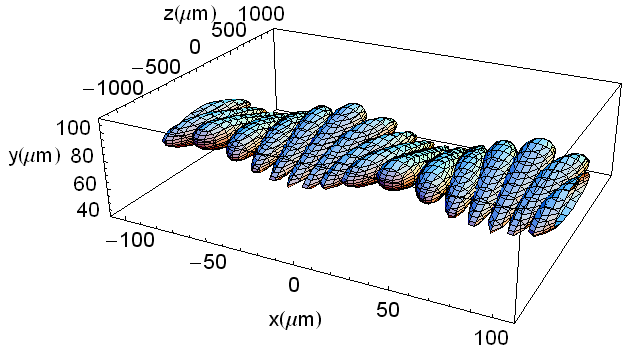}
\end{center}
\caption{\label{fig:Transport:transportSequence3D} Contour of
constant potential energy ($230\,\mu \mathrm{K}$) over two full
transport cycles. Parameters of the calculation are those used in experiments:
$\lambda=110\,\mu\mathrm{m}$, $B_b=2.2\,\mathrm{G}$, $B_{z-coil}=0$ and $I_{end}=10\,\mathrm{A}$, and gravity is included. Each contour corresponds to a different time during transport.}
\end{figure}

Figure \ref{fig:Transport:transportSequence3D} shows a 3D plot of a contour of constant potential energy (equivalent to $230\,\mu\mathrm{K}$), during a transport sequence corresponding to the same two full rotations of the bias field. As the elongated trap progresses from $x=-110\,\mu\mathrm{m}$ to $x=110\,\mu\mathrm{m}$, we observe a movement of the trap ends resembling the pedals of a bicycle, with each end describing an ellipse in the \emph{x-y} plane. The angular acceleration of the atoms at the ends of the trap due to this movement can result in transport-induced heating.\\

\begin{figure}[ht]
\begin{center}
\includegraphics[width=0.65\textwidth,height=0.87\textwidth]{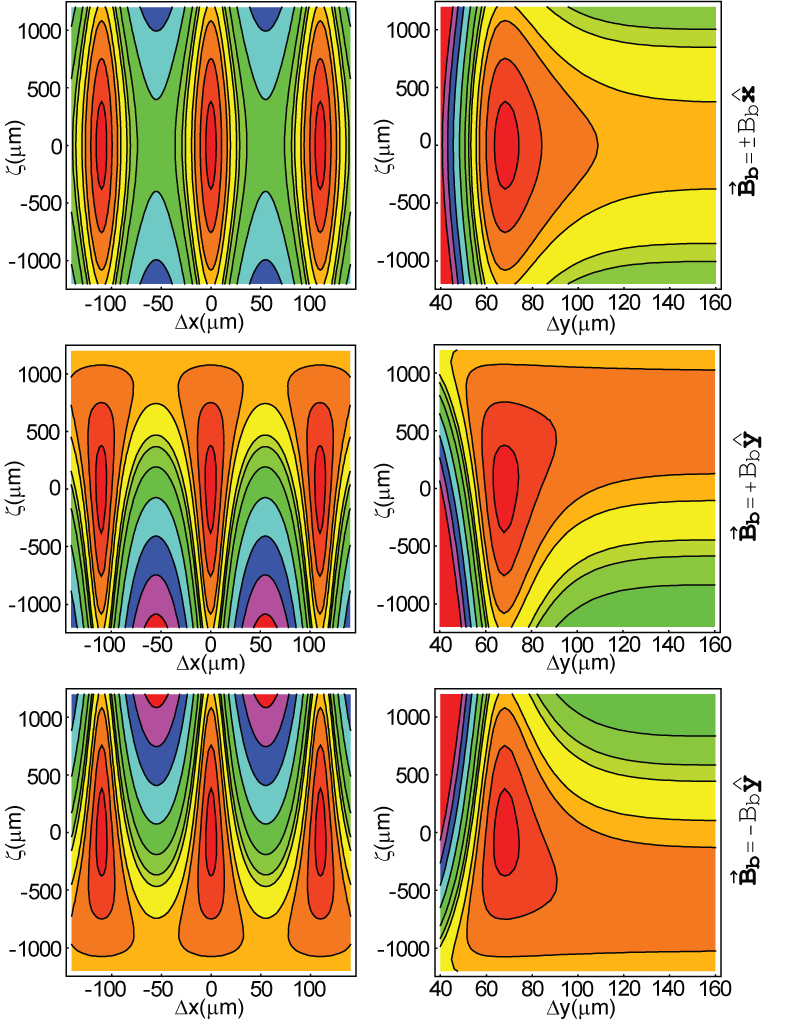}
\end{center}
\caption{\label{fig:Transport:contoursRealAxis} Contours of
constant potential energy (between $220\,\mu \mathrm{K}$ and $940\,\mu \mathrm{K}$). Cross-sections through the $x-\zeta$ and $\zeta-y$ surfaces are shown on the left- and right-hand sides respectively. $\Delta x$ and $\Delta y$ are the distances to the effective trap axis, $\zeta$, along the \emph{x} and \emph{y} directions, respectively. From top to bottom the bias fields are: $\vec{B}_b= \pm B_b \mathbf{\hat{x}}$, $\vec{B}_b=+B_b \mathbf{\hat{y}}$ and $\vec{B}_b=-B_b \mathbf{\hat{y}}$. The parameters used in these plots are: $\lambda=110\,\mu\mathrm{m}$, $B_b=2.2\,\mathrm{G}$, $B_{z-coil}=0$ and $I_{end}=10\,\mathrm{A}$. Gravity is included in the calculation.}
\end{figure}

At the positions where $\zeta$ bends vertically, there is significant variation along the trap of the transverse confinement, which is tighter at the ends closer to the videotape (due to the increased effective bias field, see equation (\ref{eqn:radialVideofreq})), and more relaxed at the opposite ends. This can be seen in figure \ref{fig:Transport:transportSequence3D}, and is further illustrated by figure \ref{fig:Transport:contoursRealAxis}, which shows the contours of constant potential energy in the videotape traps on the $x-\zeta$ and $\zeta-y$ surfaces, as we move along $\zeta$. The periodic transverse compression and de-compression of the trap ends induces an axial oscillation of the cloud (as detailed in the next section), as well as transverse excitations, heating the atoms during transport. These effects are more pronounced when the clouds being transported are hot since the change in shape of the energy equipotentials during one rotation of the bias field is more dramatic the further the atoms are from the trap axis.\\


The trap centres hardly deviate from the straight-line path along \emph{x} that we would expect for smooth transport. The \emph{y}-coordinate of the trap centre oscillates with a very small amplitude of $\sim14\,\mathrm{nm}$  at twice the bias-field-rotation frequency, while the \emph{z}-coordinate of the trap centre oscillates with an amplitude of $\sim2.2\,\mu\mathrm{m}$, at the bias-rotation frequency. The end-wire field is again responsible for these effects. The minimum potential energy calculated at the bottom of the trap stays constant during transport to within $1\,\mathrm{nK}$. These periodic variations of the trap centre during transport are very small and do not excite the atomic cloud appreciably.

\subsection{\label{sec:Transport:ExperimentalData} Experimental data}

\begin{figure}[ht]
\begin{center}
\includegraphics[width=0.5\textwidth,height=0.5\textwidth]{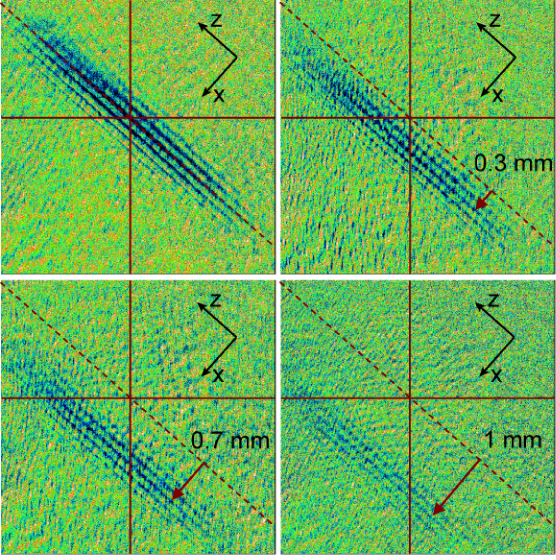}
\end{center}
\caption[Absorption images of an array of 5-6 videotape traps
transported over up to $\sim1$mm.]{\label{fig:Transport:ConveyHot}
Absorption images of an array of 5-6 videotape magnetic traps
transported parallel to the chip surface, along the \emph{x}
direction. From the top left, the
transported distances are: 0, $0.3\,\mathrm{mm}$, $0.7\,\mathrm{mm}$ and $1\,\mathrm{mm}$, corresponding to 0, 3, 6 and 9 cycles of bias field rotation at a frequency of $50\,\mathrm{Hz}$. The solid and dashed lines are fixed at the same position on all images.}
\end{figure}

Figure \ref{fig:Transport:ConveyHot} shows absorption images of
atoms confined in an array of 5-6 elongated videotape traps at temperatures around $500\,\mu \mathrm{K}$, transported up to $1\, \mathrm{mm}$ along \emph{x} as the bias field goes through 0 to 9 rotation cycles at a frequency of $50 \,\mathrm{Hz}$. The bias field is $19\, \mathrm{G}$ and the distance from the traps to the chip surface during transport is $30\,\mu \mathrm{m}$.


The obvious atom loss that we observe in figure \ref{fig:Transport:ConveyHot} is due in part to the fact that these traps are loaded without an rf-evaporation stage in the wire trap (see section \ref{sec:ExperimentalSequence}). Therefore, the ratio of videotape trap depth to atomic thermal energy is only 2, leading to an important loss rate by evaporation. Additionally, the transport-induced excitations heat the atoms, enhancing the loss. After one transport cycle we measure an induced axial dipole oscillation of the centre of mass of the cloud with an amplitude of $300\,\mu\mathrm{m}$.

When we use the optimized loading sequence described in section \ref{sec:ExperimentalSequence} and cool the atoms before conveying them, we achieve much longer transport distances. After an rf evaporation stage in the wire trap, the atoms are loaded into two videotape traps with transverse frequencies of $1.5\, \mathrm{kHz}$ and an axial frequency of $13\,\mathrm{Hz}$, at a distance of $68 \, \mu \mathrm{m}$ from the chip surface. An additional rf ramp cools the atoms down to temperatures between $6\, \mu \mathrm{K}$ and $66\, \mu \mathrm{K}$. These atoms are then transported across the surface of the chip over a distance of $\sim1\, \mathrm{cm}$ using a bias field of $2.2\, \mathrm{G}$ rotating at a frequency of $50\, \mathrm{Hz}$, to produce a transport velocity of $6\, \mathrm{mm/s}$. The magnitude of the bias field remains constant to within 2\% during the rotation.

\begin{figure}[ht]
\begin{center}
\includegraphics[width=0.5\textwidth,height=1.18\textwidth]{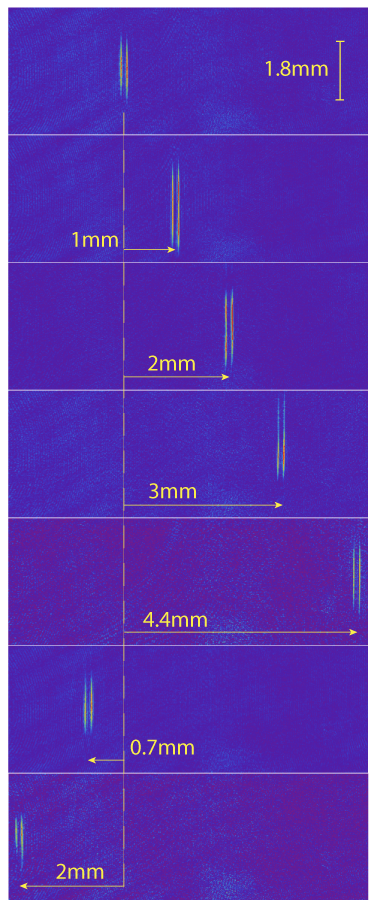}
\end{center}
\caption{\label{fig:Transport:ConveyCold} Sequences of
absorption images of cold atoms ($16\,\mu\mathrm{K}$) confined in two videotape traps and transported to either side of their initial loading position over the distances indicated. The frequency of bias field rotation is $50\,\mathrm{Hz}$, and the trap height is $68\,\mu\mathrm{m}$.}
\end{figure}

Figure \ref{fig:Transport:ConveyCold} shows a sequence of absorption images of a few times $10^5\,\mathrm{atoms}$ initially at $16\,\mu\mathrm{K}$, confined in two videotape traps and transported over $4.4\,\mathrm{mm}$ to the right, and over $2\,\mathrm{mm}$ to the left of their initial position on the image plane, corresponding to 40 and 18 full rotations of the bias-field direction, respectively. The camera is not moved during the process of data taking and the same region of interest is shown on all images. The clouds are transported until they leave the field of view on both sides of the image, with no significant atom loss.

Colder atom clouds with temperatures around $6\,\mu\mathrm{K}$ have been transported over even larger distances, up to $7\, \mathrm{mm}$ (65 bias-field-rotation cycles) towards the right, and $4\, \mathrm{mm}$ (36 cycles) towards the left-hand side, covering half the $22\,\mathrm{mm}$ length of videotape on the chip (data not shown).



\begin{figure}[ht]
\begin{center}
\includegraphics[width=0.46\textwidth,height=0.5\textwidth]{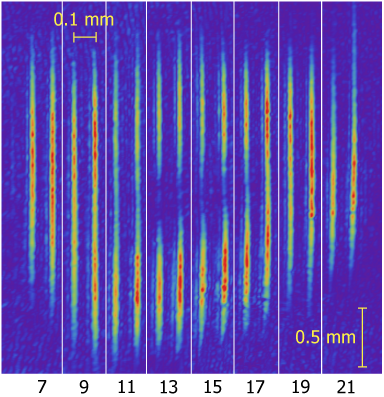}
\end{center}
\caption[Absorption images of cold atoms conveyed from 7 to 21
transport cycles.]{\label{fig:Transport:beamSplitter}
Absorption images of atoms transported from left to right, following the same experimental sequence as for the images in figure \ref{fig:Transport:ConveyCold}. Two videotape traps are initially loaded and simultaneously transported. From left to right, the clouds are conveyed over 7 to 21 transport
cycles, as indicated by the numbers below the images, corresponding
to distances between $0.8\,\mathrm{mm}$ and $2.3\,\mathrm{mm}$.
The atom-surface separation is $68\,\mu\mathrm{m}$.}
\end{figure}

We find that the inhomogeneities of the axial trapping potential change noticeably, both in shape and magnitude, as the atoms are transported to different locations away from the chip centre. In one region, after a transport distance of $\sim1.5\,\mathrm{mm}$, each atom cloud divides axially into two, separated by $\sim1.75\,\mathrm{mm}$, which
then merge again as shown by the sequence of absorption images in figure \ref{fig:Transport:beamSplitter}. Each image corresponds to a different experimental realisation in which two neighbouring clouds of atoms are conveyed over an odd number of transport cycles, from 7 to 21. The whole montage shows the clouds as they are physically positioned below the videotape. The clouds have a temperature of order $30\, \mu\mathrm{K}$. Where they split, the potential roughness has an amplitude of up to $50\,\mu\mathrm{K}$ - much more than the few $\mu$K measured in section \ref{sec:Fragmentation}.


\begin{figure}[ht]
\begin{center}
\includegraphics[width=0.258\textwidth,height=0.348\textwidth]{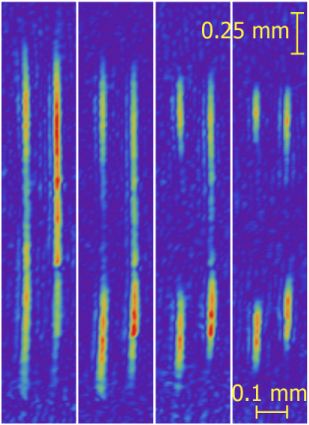}
\end{center}
\caption{\label{fig:Transport:EvapAfter18cyclesConvey}
Sequence of four absorption images of two clouds of atoms
confined in videotape traps, after they are transported over
$2\, \mathrm{mm}$ (19 cycles) and then evaporated to different final
temperatures. From left to right, the time duration of the rf sweep increases from 0 to $2\, \mathrm{s}$, and the temperature is reduced by a factor of 10, revealing the inhomogeneities of the trapping potential in this region of the videotape.}
\end{figure}

It is possible to cool the atoms by rf evaporation after having transported them. Figure \ref{fig:Transport:EvapAfter18cyclesConvey} shows a sequence of four absorption images recorded after the atoms are first conveyed over 19 transport cycles, i.e., over a distance of $2\, \mathrm{mm}$, and then cooled in the videotape traps for up to $2\, \mathrm{s}$. From left to right, the temperature of the atoms is reduced by a factor of 10. Evaporating after transport makes the potential roughness in this region more obvious, as already seen in section \ref{sec:Fragmentation}.\\


\begin{figure}[ht]
\begin{center}
\includegraphics[width=0.55\textwidth]{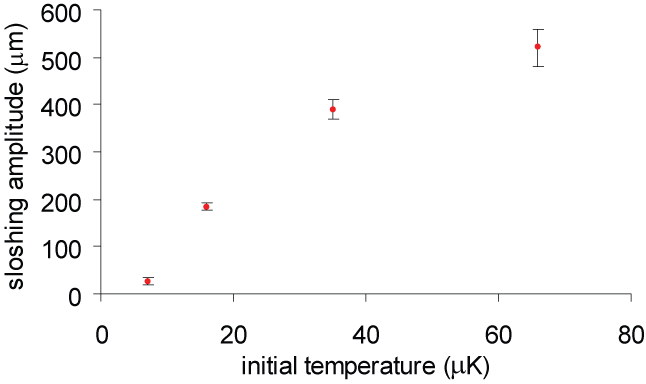}
\end{center}
\caption{\label{fig:Transport:SloshConvey} Measured amplitude of the transport-induced, axial dipole oscillations of the clouds, after one transport cycle, as a function of the initial temperature in the videotape traps before transport.}
\end{figure}

We notice that the transport process induces a clear axial excitation of the clouds. Figure \ref{fig:Transport:SloshConvey} shows the amplitude of axial oscillation of the centre of mass of the cloud, measured after one transport cycle, as a function of initial temperature. This amplitude decreases as the initial temperature of the cloud is lowered by rf evaporation before transport. This behaviour is consistent with the motion being driven by the compression and relaxation of the trap ends during transport. Indeed, a rough estimate of this type of heating captures well the observed trend of decreasing amplitude with lower temperature. The energy in this axial oscillation is responsible for heating the cloud. For example a sloshing amplitude of $400\,\mu \mathrm{m}$ dissipates to produce a temperature rise of order $2\,\mu \mathrm{K}$ per cycle, which agrees well with the heating rates we observe for clouds that are initially hot. By contrast, clouds that are initially cold have low sloshing amplitudes and are not heated substantially by this mechanism. However we do observe heating of these clouds, indicating that some other mechanism is operating as well. We have considered heating due to the initial acceleration, to the circular movement of the trap ends and to driving the cloud over the roughness of the potential. Since all of these mechanisms produce negligible heating, we presume that there is a technical cause for the measured heating that remains to be discovered. At present our heating rate is small enough to permit transport of atom clouds with temperatures of a few $\mu\mathrm{K}$ over distances of a few hundred micrometres and we see no fundamental obstacle in the future to transporting Bose-Einstein condensates in this way over long distances.

\section{\label{sec:Conclusions} Conclusions}

The results presented in this paper explain how permanently magnetized videotape mounted on an atom chip can be used for trapping, cooling and transporting ultra-cold neutral atoms.

We have demonstrated experimentally the transport of cold atoms in arrays of traps over distances as large as $1\,\mathrm{cm}$. Our transport mechanism enables us to survey the chip surface and choose a region where potential roughness is small.

We have shown that the videotape potential is smooth enough to allow a wide range of experiments with cold atoms, and that the potential roughness is quite low compared with other atom-chip experiments (with the exception of reference \cite{f17}). Furthermore, we have proposed a method to reduce fragmentation in our traps, which could make our chip suitable for studies of one-dimensional quantum gases.

\ack
We thank Professor Horst D\"{o}tsch, from Osnabr\"{u}ck University in Germany, for providing us with the garnet sensor that allowed us to image and measure the videotape magnetisation. We are indebted to Manuel Succo for setting up the rf generator used in the evaporation. This work was funded by the FastNet and AtomChips European networks and the UK EPSRC and Royal Society.

\section*{References}
\providecommand{\newblock}{}


\begin{thebibliography}{10}
\expandafter\ifx\csname url\endcsname\relax
  \def\url#1{{\tt #1}}\fi
\expandafter\ifx\csname urlprefix\endcsname\relax\def\urlprefix{URL }\fi
\providecommand{\eprint}[2][]{\url{#2}}

\bibitem{chips9}
Reichel J, H\"{a}nsel W and H\"{a}nsch T~W 1999 {\em Phys. Rev. Lett.\/} {\bf
  83} 3398

\bibitem{t15}
Folman R, Kr\"{u}ger P, Cassettari D, Hessmo B, Maier T and Schmiedmayer J 2000
  {\em Phys. Rev. Lett.\/} {\bf 84} 4749

\bibitem{intf1}
Schumm T, Hofferberth S, Andersson L~M, Wildermuth S, Groth S, Bar-Joseph I,
  Schmiedmayer J and Kr\"{u}ger P 2005 {\em Nature Physics\/} {\bf 1} 57

\bibitem{intf3}
G\"{u}nther A, Kraft S, Zimmermann C and Fort\'{a}gh J 2007 {\em Phys. Rev.
  Lett.\/} {\bf 98} 140403

\bibitem{intf5}
Jo G~B, Choi J~H, Christensen C~A, Lee Y~R, Pasquini T~A, Ketterle W and
  Pritchard D~E 2007 {\em Phys. Rev. Lett.\/} {\bf 99} 240406

\bibitem{1DonChip1}
Esteve J, Trebbia J~B, Schumm T, Aspect A, Westbrook C~I and Bouchoule I 2006
  {\em Phys. Rev. Lett.\/} {\bf 96} 130403

\bibitem{1DonChip2}
Trebbia J~B, Esteve J, Westbrook C~I and Bouchoule I 2006 {\em Phys. Rev.
  Lett.\/} {\bf 97} 250403

\bibitem{1DonChip3}
Hofferberth S, Lesanovsky I, Fischer B, Schumm T and Schmiedmayer J 2007 {\em
  Nature\/} {\bf 449} 324

\bibitem{1DonChip4}
{van Amerongen} A~H, {van Es} J~J~P, Wicke P, Kheruntsyan K~V and {van Druten}
  N~J 2008 {\em Phys. Rev. Lett.\/} {\bf 100} 090402

\bibitem{intf6}
Hofferberth S, Lesanovsky I, Schumm T, Imambekov A, Gritsev V, Demler E and
  Schmiedmayer J 2008 {\em Nature Physics\/} {\bf 4} 489

\bibitem{JonesSpinFlips}
Jones M~P~A, Vale C~J, Sahagun D, Hall B~V and Hinds E~A 2003 {\em Phys. Rev.
  Lett.\/} {\bf 91} 080401

\bibitem{CasimirPolder3}
Lin Y, Teper I, Chin C and Vuleti\'{c} V 2004 {\em Phys. Rev. Lett.\/} {\bf 92}
  050404

\bibitem{ThermalSpinFlips}
Scheel S, Rekdal P~K, Knight P~L and Hinds E~A 2005 {\em Phys. Rev. A\/} {\bf
  72} 042901

\bibitem{f14}
Wildermuth S, Hofferberth S, Lesanovsky I, Haller E, Andersson L~M, Groth S,
  Bar-Joseph I, Kr\"{u}ger P and Schmiedmayer J 2005 {\em Nature\/} {\bf 435}
  440

\bibitem{f16}
Wildermuth S, Hofferberth S, Lesanovsky I, Groth S, Kr\"{u}ger P, Schmiedmayer
  J and Bar-Joseph I 2006 {\em Appl. Phys. Lett.\/} {\bf 88} 264103

\bibitem{f26}
Aigner S, Pietra L~D, Japha Y, Entin-Wohlman O, David T, Salem R, Folman R and
  Schmiedmayer J 2008 {\em Science\/} {\bf 319} 226

\bibitem{VortexSensor}
Scheel S, Fermani R and Hinds E~A 2007 {\em Phys. Rev. A\/} {\bf 75} 064901

\bibitem{AtomicClockChip1}
Knappe S, Schwindt P~D~D, Shah V, Hollberg L, Kitching J, Liew L and Moreland J
  2005 {\em Optics Express\/} {\bf 13} 1249

\bibitem{AtomicClockChipFernando}
Lacro\^{u}te C, Reinhard F, Ramirez-Martinez F, Deutsch C, Schneider T, Reichel
  J and Rosenbusch P 2010 {\em IEEE Transactions on Ultrasonics, Ferroelectrics
  and Frequency Control\/} {\bf 57} 106

\bibitem{qi2}
Treutlein P, Steinmetz T, Colombe Y, Hommelhoff P, Reichel J, Greiner M, Mandel
  O, Widera A, Rom T, Bloch I and H\"{a}nsch T~W 2006 {\em Fortschr. Phys.\/}
  {\bf 54} 702

\bibitem{qi3}
Trupke M, Metz J, Beige A and Hinds E~A 2007 {\em Journal of Modern Optics\/}
  {\bf 54} 1639

\bibitem{qi4}
Kr\"{u}ger P, Haase A, Andersson L~M and Schmiedmayer J 2002 {\em Journal of
  Modern Optics\/} {\bf 49} 1375

\bibitem{Detection5}
Takamizawa A, Steinmetz T, Delhuille R, H\"{a}nsch T~W and Reichel J 2006 {\em
  Optics Express\/} {\bf 14} 10976

\bibitem{Detection11}
Colombe Y, Steinmetz T, Dubois G, Linke F, Hunger D and Reichel J 2007 {\em
  Nature\/} {\bf 450} 272

\bibitem{DetectionCavity4}
Teper I, Lin Y~J and Vuleti\'{c} V 2006 {\em Phys. Rev. Lett.\/} {\bf 97}
  023002

\bibitem{Detection10}
Purdy T~P and Stamper-Kurn D~M 2008 {\em Appl. Phys. B\/} {\bf 90} 401

\bibitem{Detection8}
Heine D, Wilzbach M, Raub T, Hessmo B and Schmiedmayer J 2009 {\em Phys. Rev.
  A\/} {\bf 79} 021804(R)

\bibitem{DetectionCavity5}
Trupke M, Goldwin J, Darqui\'{e} B, Dutier G, Eriksson S, Ashmore J and Hinds
  E~A 2007 {\em Phys. Rev. Lett.\/} {\bf 99} 063601

\bibitem{pyramids}
Pollock S, Cotter J~P, Laliotis A and Hinds E~A 2009 {\em Opt. Express\/} {\bf
  17} 14109

\bibitem{waveguidesPaper}
Kohnen M, Succo M, Petrov P~G, Nyman R~A, Trupke M and Hinds E~A 2010 {\em
  arXiv:0912.4460\/}

\bibitem{chipsReview3}
Fort\'{a}gh J and Zimmermann C 2007 {\em Rev. Mod. Phys.\/} {\bf 79} 235

\bibitem{PermanentMagnets1}
Hall B~V, Whitlock S, Scharnberg F, Hannaford P and Sidorov A 2006 {\em J.
  Phys. B: At. Mol. Opt. Phys.\/} {\bf 39} 27

\bibitem{PermanentMagnets3}
Singh M, Volk M, Akulshin A, Sidorov A, McLean R and Hannaford P 2008 {\em J.
  Phys. B: At. Mol. Opt. Phys.\/} {\bf 41} 065301

\bibitem{PermanentMagnets0}
Barb I, Gerritsma R, Xing Y~T, Goedkoop J~B and Spreeuw R~J~C 2005 {\em Eur.
  Phys. J. D\/} {\bf 35} 75

\bibitem{PermanentMagnets6}
Gerritsma R, Whitlock S, Fernholz T, Schlatter H, Luigjes J~A, Thiele J~U,
  Goedkoop J~B and Spreeuw R~J~C 2007 {\em Phys. Rev. A\/} {\bf 76} 033408

\bibitem{PermanentMagnets7}
Fernholz T, Gerritsma R, Whitlock S, Barb I and Spreeuw R~J~C 2008 {\em Phys.
  Rev. A\/} {\bf 77} 033409

\bibitem{f22}
Boyd M, Streed E~W, Medley P, Campbell G~K, Mun J, Ketterle W and Pritchard D~E
  2007 {\em Phys. Rev. A\/} {\bf 76} 043624

\bibitem{PermanentMagnets4}
Jaakkola A, Shevchenko A, Lindfors K, Hautakorpi M, Il'yashenkoa E, Johansen
  T~H and Kaivola M 2005 {\em Eur. Phys. J. D\/} {\bf 35} 81

\bibitem{PermanentMagnets5}
Shevchenko A, Heili\"{o} M, Lindvall T, Jaakkola A, Tittonen I, Kaivola M and
  Pfau T 2006 {\em Phys. Rev. A\/} {\bf 73} 051401(R)

\bibitem{chips5}
Hughes I~G, Barton P~A, Roach T~M, Boshier M~G and Hinds E~A 1997 {\em J. Phys.
  B: At. Mol. Opt. Phys.\/} {\bf 30} 647

\bibitem{videofieldcalculation}
Hughes I~G, Barton P~A, Roach T~M and Hinds E~A 1997 {\em J. Phys. B: At. Mol.
  Opt. Phys.\/} {\bf 30} 2119

\bibitem{chips7}
Saba C~V, Barton P~A, Boshier M~G, Hughes I~G, Rosenbusch P, Sauer B~E and
  Hinds E~A 1999 {\em Phys. Rev. Lett.\/} {\bf 82} 468

\bibitem{chips8}
Rosenbusch P, Hall B~V, Hughes I~G, Saba C~V and Hinds E~A 2000 {\em Appl.
  Phys. B\/} {\bf 70} 709

\bibitem{t16}
Hinds E~A and Hughes I~G 1999 {\em J. Phys. D: Appl. Phys.\/} {\bf 32} R119

\bibitem{t22}
Hinds E~A 1999 {\em Phil. Trans. R. Soc. Lond. A\/} {\bf 357} 1409

\bibitem{PermanentMagnets2}
Ghanbari S, Kieu T~D and Hannaford P 2007 {\em J. Phys. B: At. Mol. Opt.
  Phys.\/} {\bf 40} 1283

\bibitem{mofilms}
Eriksson S, Ramirez-Martinez F, Curtis E~A, Sauer B~E, Nutter P~W, Hill E~W and
  Hinds E~A 2004 {\em Appl. Phys. B\/} {\bf 79} 811

\bibitem{ChrisPRA}
Sinclair C~D~J, Curtis E~A, {Llorente Garc\'{i}a} I, Retter J~A, Hall B~V,
  Eriksson S, Sauer B~E and Hinds E~A 2005 {\em Phys. Rev. A\/} {\bf 72}
  {031603(R)}

\bibitem{ChrisEPJD}
Sinclair C~D~J, Retter J~A, Curtis E~A, Hall B~V, {Llorente Garc\'{i}a} I,
  Eriksson S, Sauer B~E and Hinds E~A 2005 {\em Eur. Phys. J. D\/} {\bf 35} 105

\bibitem{Doetsch}
Klank M, Hagedorn O, Holthaus C, Shamonin M and D\"{o}tsch H 2003 {\em {NDT \&
  E International}\/} {\bf 36} 375

\bibitem{MyThesis}
{Llorente Garc\'{i}a} I 2008 {\em {Advances in the design and operation of atom
  chips}\/} Ph.D. thesis {Imperial College London}

\bibitem{LVIS}
Lu Z~T, Corwin K~L, Renn M~J, Anderson M~H, Cornell E~A and Wieman C~E 1996
  {\em Phys. Rev. Lett.\/} {\bf 77} 3331

\bibitem{interferometerSewell}
Sewell R~J, Dingjan J, Baumg\"{a}rtner F, Llorente-Garc\'{\i}a I, Eriksson S,
  Hinds E~A, Lewis G, Srinivasan P, Moktadir Z, Gollasch C~O and Kraft M 2010
  {\em J. Phys. B: At. Mol. Opt. Phys.\/} {\bf 43} 051003

\bibitem{f21}
Hall B~V, Whitlock S, Anderson R, Hannaford P and Sidorov A~I 2007 {\em Phys.
  Rev. Lett.\/} {\bf 98} 030402

\bibitem{f1}
Fort\'{a}gh J, Ott H, Kraft S, G\"{u}nther A and Zimmermann C 2002 {\em Phys.
  Rev. A\/} {\bf 66} 041604(R)

\bibitem{f2}
Kraft S, G\"{u}nther A, Ott H, Wharam D, Zimmermann C and Fort\'{a}gh J 2002
  {\em J. Phys. B: At. Mol. Opt. Phys.\/} {\bf 35} L469

\bibitem{f3}
Fort\'{a}gh J, Ott H, Kraft S, G\"{u}nther A and Zimmermann C 2003 {\em Appl.
  Phys. B\/} {\bf 76} 157

\bibitem{f4}
Leanhardt A~E, Chikkatur A~P, Kielpinski D, Shin Y, Gustavson T~L, Ketterle W
  and Pritchard D~E 2002 {\em Phys. Rev. Lett.\/} {\bf 89} 040401

\bibitem{f7}
Leanhardt A~E, Shin Y, Chikkatur A~P, Kielpinski D, Ketterle W and Pritchard
  D~E 2003 {\em Phys. Rev. Lett.\/} {\bf 90} 100404

\bibitem{f6}
Jones M~P~A, Vale C~J, Sahagun D, Hall B~V and Hinds E~A 2003 {\em Phys. Rev.
  Lett.\/} {\bf 91} 080401

\bibitem{f5}
Jones M~P~A, Vale C~J, Sahagun D, Hall B~V, Eberlein C~C, Sauer B~E, Furusawa
  K, Richardson D and Hinds E~A 2004 {\em J. Phys. B: At. Mol. Opt. Phys.\/}
  {\bf 37} L15

\bibitem{f10}
Vale C~J, Upcroft B, Davis M~J, Heckenberg N~R and Rubinsztein-Dunlop H 2004
  {\em J. Phys. B: At. Mol. Opt. Phys.\/} {\bf 37} 2959

\bibitem{f8}
Est\`{e}ve J, Aussibal C, Schumm T, Figl C, Mailly D, Bouchoule I, Westbrook
  C~I and Aspect A 2004 {\em Phys. Rev. A\/} {\bf 70} 043629

\bibitem{f9b}
Wang D~W, Lukin M~D and Demler E 2004 {\em Phys. Rev. Lett.\/} {\bf 92} 076802

\bibitem{f11}
Schumm T, Est\`{e}ve J, Figl C, Trebbia J~B, Aussibal C, Nguyen H, Mailly D,
  Bouchoule I, Westbrook C~I and Aspect A 2005 {\em Eur. Phys. J. D\/} {\bf 32}
  171

\bibitem{f17}
Kr\"{u}ger P, Andersson L~M, Wildermuth S, Hofferberth S, Haller E, Aigner S,
  Groth S, Bar-Joseph I and Schmiedmayer J 2007 {\em Phys. Rev. A\/} {\bf 76}
  063621

\bibitem{f36}
Moktadir Z, Darqui\'{e} B, Kraft M and Hinds E~A 2007 {\em Journal of Modern
  Optics\/} {\bf 54} 2149

\bibitem{f19}
Whitlock S, Hall B~V, Roach T, Anderson R, Volk M, Hannaford P and Sidorov A~I
  2007 {\em Phys. Rev. A\/} {\bf 75} 043602

\bibitem{ChrisThesis}
Sinclair C~D~J 2005 {\em {Bose-Einstein Condensation in Microtraps on
  Videotape}\/} Ph.D. thesis {Imperial College London}

\bibitem{f24}
Trebbia J~B, {Garrido Alzar} C~L, Cornelussen R, Westbrook C~I and Bouchoule I
  2007 {\em Phys. Rev. Lett.\/} {\bf 98} 263201

\bibitem{st18}
Bouchoule I, Trebbia J~B and {Garrido Alzar} C~L 2008 {\em Phys. Rev. A\/} {\bf
  77} 023624

\bibitem{TOPtrap}
Petrich W, Anderson M~H, Ensher J~R and Cornell E~A 1995 {\em Phys. Rev.
  Lett.\/} {\bf 74} 3352

\bibitem{TonksReichelBoxLike}
Reichel J and Thywissen J~H 2004 {\em J. Phys. IV France\/} {\bf 116} 265

\bibitem{t19}
Dekker N~H, Lee C~S, Lorent V, Thywissen J~H, Smith S~P, Drnd\'{i}c M,
  Westervelt R~M and Prentiss M 2000 {\em Phys. Rev. Lett.\/} {\bf 84} 1124

\bibitem{t8}
H\"{a}nsel W, Reichel J, Hommelhoff P and H\"{a}nsch T~W 2001 {\em Phys. Rev.
  Lett.\/} {\bf 86} 608

\bibitem{t2}
H\"{a}nsel W, Hommelhoff P, H\"{a}nsch T~W and Reichel J 2001 {\em Nature\/}
  {\bf 413} 498

\bibitem{t1}
Hommelhoff P, H\"{a}nsel W, Steinmetz T, H\"{a}nsch T~W and Reichel J 2005 {\em
  New Journal of Physics\/} {\bf 7} 063621

\bibitem{t33}
Long R, Rom T, H\"{a}nsel W, H\"{a}nsch T~W and Reichel J 2005 {\em Eur. Phys.
  J. D\/} {\bf 35} 125

\bibitem{t10}
G\"{u}nther A, Kemmler M, Kraft S, Vale C~J, Zimmermann C and Fort\'{a}gh J
  2005 {\em Phys. Rev. A\/} {\bf 71} 063619

\bibitem{t20}
Whitlock S, Gerritsma R, Fernholz T and Spreeuw R~J~C 2009 {\em New J. Phys.\/}
  {\bf 11} 023021

\bibitem{t18}
Rosenbusch P, Hall B~V, Hughes I~G, Saba C~V and Hinds E~A 2000 {\em Phys. Rev.
  A\/} {\bf 61} 031404(R)

\end{thebibliography}
\end{document}